\documentclass[aps,prd,reprint,preprintnumbers,showpacs,nofootinbib,superscript address,longbibliography,floatfix]{revtex4-2}
\usepackage[utf8]{inputenc}
\usepackage{amsmath,amsfonts,amssymb,amsfonts}
\usepackage{fontawesome}
\usepackage{pifont}
\usepackage{mathrsfs}
\usepackage{accents}
\usepackage{verbatim}
\usepackage{natbib}
\usepackage[english]{babel}
\usepackage{adjustbox}
\usepackage{relsize} 
\usepackage{slashed}
\usepackage{scalerel,stackengine}
\usepackage{hhline}
\usepackage{mathtools}
\usepackage[dvipsnames]{xcolor}
\usepackage{longtable}
\usepackage{array}
\usepackage{siunitx}
\usepackage{graphicx}
\usepackage{xstring}
\usepackage{etoolbox}
\usepackage{notoccite}
\usepackage{tensor}
\usepackage{tocbasic}
\DeclareTOCStyleEntry[numwidth=25pt,linefill=\bfseries\TOCLineLeaderFill]{tocline}{section}
\DeclareTOCStyleEntry[entryformat=\textit,numwidth=10pt,linefill=\TOCLineLeaderFill]{tocline}{subsection}
\DeclareTOCStyleEntry[entryformat=\textit,numwidth=10pt,linefill=\TOCLineLeaderFill]{tocline}{subsubsection}
\usepackage{xspace}

\usepackage{tikz}
\usetikzlibrary{calc}
\usepackage{pgfkeys}
\pgfdeclarelayer{foreground}
\pgfsetlayers{main,foreground}

\newrobustcmd{\PSALTer}{\textit{PSALTer}\xspace}

\newcommand{\SampleConsistent}{\color{SeaGreen}\faCheck\color{black}}%
\newcommand{\SampleInconsistent}{\color{red}\faRemove\color{black}}%
\newcommand{\SampleEmpty}{\color{gray}\faCircleO\color{black}}%

\newcommand{\Consistent}{\Large\color{SeaGreen}\faCheck\color{black}}%
\newcommand{\Inconsistent}{\Large\color{red}\faRemove\color{black}}%
\newcommand{\Empty}{\Large\color{gray}\faCircleO\color{black}}%

\newrobustcmd{\GenXExpr}[1]{%
	\tensor[^{(#1)}]{X}{}%
}
\newrobustcmd{\GenLExpr}[1]{%
	\tensor[^{(#1)}]{L}{}%
}
\newrobustcmd{\GenNExpr}[1]{%
	\tensor[^{(#1)}]{N}{}%
}

\newrobustcmd{\GenNullVector}[1]{%
	\tensor[^{(#1)}]{\mathsf{v}}{}%
}

\newrobustcmd{\XExpr}[3]{%
	{\tensor[^{({#3})}]{X}{_{{{#1}^{#2}}}}}%
}
\newrobustcmd{\LExpr}[3]{%
	{\tensor[^{({#3})}]{L}{_{{{#1}^{#2}}}}}%
}
\newrobustcmd{\NExpr}[3]{%
	{\tensor[^{({#3})}]{N}{_{{{#1}^{#2}}}}}%
}
\newrobustcmd{\SubXExpr}[4]{%
	{\tensor*[^{({#3})}_{({#4})}]{X}{_{{{#1}^{#2}}}}}%
}
\newrobustcmd{\SubLExpr}[4]{%
	{\tensor*[^{({#3})}_{({#4})}]{L}{_{{{#1}^{#2}}}}}%
}
\newrobustcmd{\SubNExpr}[4]{%
	{\tensor*[^{({#3})}_{({#4})}]{N}{_{{{#1}^{#2}}}}}%
}
\newrobustcmd{\NullVector}[3]{%
	{\tensor[^{(#3)}]{\mathsf{v}}{_{{{#1}^{#2}}}}}%
}

\newrobustcmd{\Dis}[1]{%
	\tensor{\mathcal{K}}{#1}%
}
\newrobustcmd{\Cur}[1]{%
	{\tensor{\mathcal{J}}{#1}}%
}
\newrobustcmd{\OperatorO}[2]{%
	\tensor*{\mathcal{O}}{^{{#1}}_{(#2)}}
}
\newrobustcmd{\hel}[2][]{%
  \ifstrempty{#1}{%
    \tensor*{\varepsilon}{_{#2}}%
  }{%
    \tensor*{\varepsilon}{_{#2}^{\##1}}%
  }%
}
\newrobustcmd{\MAGG}[2]{%
	\tensor{\vphantom{l^{l^2}}\smash{\stackrel{\raisebox{-3pt}{\scalebox{0.55}{$({#1})$}}}{\scalebox{0.99}{$\kappa$}}}}{#2}%
}
\newrobustcmd{\PD}[1]{%
	\tensor{\partial}{#1}%
}
\newrobustcmd{\LorentzGroup}{%
	\mathrm{SO}^+(3,1)
}

\newrobustcmd{\NamedModel}[1]{\ifstrequal{#1}{A23}{$\aleph$}{\ifstrequal{#1}{A23B1}{$\tensor*{{\mathfrak{B}}}{_{1}}$}{\ifstrequal{#1}{A23B1D1}{$\tensor*{{\mathfrak{D}}}{_{1}}$}{\ifstrequal{#1}{A23B1D1E1}{$\tensor*{{\mathfrak{E}}}{_{14}}$}{\ifstrequal{#1}{A23B1D1E1G1}{$\tensor*{{\mathfrak{G}}}{_{21}}$}{\ifstrequal{#1}{A23B1D1E1G2}{$\tensor*{{\mathfrak{G}}}{_{22}}$}{\ifstrequal{#1}{A23B1D1E1G2H1}{$\tensor*{{\mathfrak{H}}}{_{23}}$}{\ifstrequal{#1}{A23B1D1E1G2H1I1}{$\tensor*{{\mathfrak{I}}}{_{28}}$}{\ifstrequal{#1}{A23B1D1E1G2H1I1K1}{$\tensor*{{\mathfrak{K}}}{_{1}}$}{\ifstrequal{#1}{A23B1D1E1G2H1I1K2}{$\tensor*{{\mathfrak{K}}}{_{2}}$}{\ifstrequal{#1}{A23B1D1E1G2H1J1}{$\tensor*{{\mathfrak{J}}}{_{11}}$}{\ifstrequal{#1}{A23B1D1E1G2H1J1K1}{$\tensor*{{\mathfrak{K}}}{_{3}}$}{\ifstrequal{#1}{A23B1D1E1G2H1J2}{$\tensor*{{\mathfrak{J}}}{_{12}}$}{\ifstrequal{#1}{A23B1D1E1G2I1}{$\tensor*{{\mathfrak{I}}}{_{24}}$}{\ifstrequal{#1}{A23B1D1E1G2I1J1}{$\tensor*{{\mathfrak{J}}}{_{28}}$}{\ifstrequal{#1}{A23B1D1E1G2I2}{$\tensor*{{\mathfrak{I}}}{_{25}}$}{\ifstrequal{#1}{A23B1D1F1}{$\tensor*{{\mathfrak{F}}}{_{4}}$}{\ifstrequal{#1}{A23B1D1F2}{$\tensor*{{\mathfrak{F}}}{_{5}}$}{\ifstrequal{#1}{A23B1D1F2H1}{$\tensor*{{\mathfrak{H}}}{_{7}}$}{\ifstrequal{#1}{A23B1D1F2H1I2}{$\tensor*{{\mathfrak{I}}}{_{33}}$}{\ifstrequal{#1}{A23B1D1F2H2}{$\tensor*{{\mathfrak{H}}}{_{8}}$}{\ifstrequal{#1}{A23B1D2}{$\tensor*{{\mathfrak{D}}}{_{2}}$}{\ifstrequal{#1}{A23B1D2F1}{$\tensor*{{\mathfrak{F}}}{_{6}}$}{\ifstrequal{#1}{A23B1D2F2}{$\tensor*{{\mathfrak{F}}}{_{7}}$}{\ifstrequal{#1}{A23B1D2F2H1}{$\tensor*{{\mathfrak{H}}}{_{9}}$}{\ifstrequal{#1}{A23B1D2F2H1I2}{$\tensor*{{\mathfrak{I}}}{_{34}}$}{\ifstrequal{#1}{A23B1D2F2H2}{$\tensor*{{\mathfrak{H}}}{_{10}}$}{\ifstrequal{#1}{A23B1D3}{$\tensor*{{\mathfrak{D}}}{_{3}}$}{\ifstrequal{#1}{A23B1D3F3}{$\tensor*{{\mathfrak{F}}}{_{8}}$}{\ifstrequal{#1}{A23B1D3F3H1}{$\tensor*{{\mathfrak{H}}}{_{11}}$}{\ifstrequal{#1}{A23B1D3F3H1I1}{$\tensor*{{\mathfrak{I}}}{_{30}}$}{\ifstrequal{#1}{A23B1D3F3H1I1J1}{$\tensor*{{\mathfrak{J}}}{_{27}}$}{\ifstrequal{#1}{A23B1D3F3H1I2}{$\tensor*{{\mathfrak{I}}}{_{29}}$}{\ifstrequal{#1}{A23B1D3F3H2}{$\tensor*{{\mathfrak{H}}}{_{12}}$}{\ifstrequal{#1}{A23B1D3F3H2I2}{$\tensor*{{\mathfrak{I}}}{_{31}}$}{\ifstrequal{#1}{A23B1D3F3H3}{$\tensor*{{\mathfrak{H}}}{_{13}}$}{\ifstrequal{#1}{A23B1D3F3H4}{$\tensor*{{\mathfrak{H}}}{_{14}}$}{\ifstrequal{#1}{A23B1D3F3H4J1}{$\tensor*{{\mathfrak{J}}}{_{1}}$}{\ifstrequal{#1}{A23B1D3F3H4J1K1}{$\tensor*{{\mathfrak{K}}}{_{5}}$}{\ifstrequal{#1}{A23B1D3F3H4J1K2}{$\tensor*{{\mathfrak{K}}}{_{6}}$}{\ifstrequal{#1}{A23B1D3F3H4J1K3}{$\tensor*{{\mathfrak{K}}}{_{7}}$}{\ifstrequal{#1}{A23B1D3F3H4J2}{$\tensor*{{\mathfrak{J}}}{_{2}}$}{\ifstrequal{#1}{A23B1D3F3H4J2K2}{$\tensor*{{\mathfrak{K}}}{_{8}}$}{\ifstrequal{#1}{A23B1D3F3H4J2K3}{$\tensor*{{\mathfrak{K}}}{_{9}}$}{\ifstrequal{#1}{A23B1D3F3H4J3}{$\tensor*{{\mathfrak{J}}}{_{3}}$}{\ifstrequal{#1}{A23B1D3F3H4J3K3}{$\tensor*{{\mathfrak{K}}}{_{10}}$}{\ifstrequal{#1}{A23B1D3F3H4J4}{$\tensor*{{\mathfrak{J}}}{_{4}}$}{\ifstrequal{#1}{A23B1D3F4}{$\tensor*{{\mathfrak{F}}}{_{9}}$}{\ifstrequal{#1}{A23B1D3F4H1}{$\tensor*{{\mathfrak{H}}}{_{15}}$}{\ifstrequal{#1}{A23B1D3F4H1I2}{$\tensor*{{\mathfrak{I}}}{_{35}}$}{\ifstrequal{#1}{A23B1D3F4H2}{$\tensor*{{\mathfrak{H}}}{_{16}}$}{\ifstrequal{#1}{A23B1D3F4H3}{$\tensor*{{\mathfrak{H}}}{_{17}}$}{\ifstrequal{#1}{A23B1D4}{$\tensor*{{\mathfrak{D}}}{_{4}}$}{\ifstrequal{#1}{A23B1D4F1}{$\tensor*{{\mathfrak{F}}}{_{10}}$}{\ifstrequal{#1}{A23B1D4F1G1}{$\tensor*{{\mathfrak{G}}}{_{28}}$}{\ifstrequal{#1}{A23B1D4F1G1H1}{$\tensor*{{\mathfrak{H}}}{_{34}}$}{\ifstrequal{#1}{A23B1D4F1G2}{$\tensor*{{\mathfrak{G}}}{_{29}}$}{\ifstrequal{#1}{A23B1D4F1H2}{$\tensor*{{\mathfrak{H}}}{_{18}}$}{\ifstrequal{#1}{A23B1D4F1H2J1}{$\tensor*{{\mathfrak{J}}}{_{5}}$}{\ifstrequal{#1}{A23B1D4F1H2J2}{$\tensor*{{\mathfrak{J}}}{_{6}}$}{\ifstrequal{#1}{A23B1D4F2}{$\tensor*{{\mathfrak{F}}}{_{11}}$}{\ifstrequal{#1}{A23B1D4F2G2}{$\tensor*{{\mathfrak{G}}}{_{30}}$}{\ifstrequal{#1}{A23B1D4F2H1}{$\tensor*{{\mathfrak{H}}}{_{19}}$}{\ifstrequal{#1}{A23B1D4F2H1J1}{$\tensor*{{\mathfrak{J}}}{_{7}}$}{\ifstrequal{#1}{A23B1D4F2H1J2}{$\tensor*{{\mathfrak{J}}}{_{8}}$}{\ifstrequal{#1}{A23B1D4F3}{$\tensor*{{\mathfrak{F}}}{_{12}}$}{\ifstrequal{#1}{A23B1D4F3H2}{$\tensor*{{\mathfrak{H}}}{_{20}}$}{\ifstrequal{#1}{A23B1D4F3H2J1}{$\tensor*{{\mathfrak{J}}}{_{9}}$}{\ifstrequal{#1}{A23B1D4F3H2J2}{$\tensor*{{\mathfrak{J}}}{_{10}}$}{\ifstrequal{#1}{A23B1D4G1}{$\tensor*{{\mathfrak{G}}}{_{1}}$}{\ifstrequal{#1}{A23B1D4G1I1}{$\tensor*{{\mathfrak{I}}}{_{1}}$}{\ifstrequal{#1}{A23B1D4G1I2}{$\tensor*{{\mathfrak{I}}}{_{2}}$}{\ifstrequal{#1}{A23B1D4G2}{$\tensor*{{\mathfrak{G}}}{_{2}}$}{\ifstrequal{#1}{A23B1D4G2I1}{$\tensor*{{\mathfrak{I}}}{_{3}}$}{\ifstrequal{#1}{A23B1D4G2I1J2}{$\tensor*{{\mathfrak{J}}}{_{21}}$}{\ifstrequal{#1}{A23B1D4G2I1J3}{$\tensor*{{\mathfrak{J}}}{_{22}}$}{\ifstrequal{#1}{A23B1D4G2I3}{$\tensor*{{\mathfrak{I}}}{_{4}}$}{\ifstrequal{#1}{A23B1D4G2I3J3}{$\tensor*{{\mathfrak{J}}}{_{23}}$}{\ifstrequal{#1}{A23B1D4G2I4}{$\tensor*{{\mathfrak{I}}}{_{5}}$}{\ifstrequal{#1}{A23B1E1}{$\tensor*{{\mathfrak{E}}}{_{1}}$}{\ifstrequal{#1}{A23B1E1G1}{$\tensor*{{\mathfrak{G}}}{_{5}}$}{\ifstrequal{#1}{A23B1E1G2}{$\tensor*{{\mathfrak{G}}}{_{6}}$}{\ifstrequal{#1}{A23B1E2}{$\tensor*{{\mathfrak{E}}}{_{2}}$}{\ifstrequal{#1}{A23B1E2G1}{$\tensor*{{\mathfrak{G}}}{_{7}}$}{\ifstrequal{#1}{A23B1E2G1H2}{$\tensor*{{\mathfrak{H}}}{_{33}}$}{\ifstrequal{#1}{A23B1E2G4}{$\tensor*{{\mathfrak{G}}}{_{8}}$}{\ifstrequal{#1}{A23B2}{$\tensor*{{\mathfrak{B}}}{_{2}}$}{\ifstrequal{#1}{A23B2C1}{$\tensor*{{\mathfrak{C}}}{_{5}}$}{\ifstrequal{#1}{A23B2C1E2}{$\tensor*{{\mathfrak{E}}}{_{15}}$}{\ifstrequal{#1}{A23B2C1E2G2}{$\tensor*{{\mathfrak{G}}}{_{23}}$}{\ifstrequal{#1}{A23B2C1E2G2H1}{$\tensor*{{\mathfrak{H}}}{_{26}}$}{\ifstrequal{#1}{A23B2C1E2G2H1I1}{$\tensor*{{\mathfrak{I}}}{_{32}}$}{\ifstrequal{#1}{A23B2C1E2G2H1I1K2}{$\tensor*{{\mathfrak{K}}}{_{4}}$}{\ifstrequal{#1}{A23B2C1E2G2H1J2}{$\tensor*{{\mathfrak{J}}}{_{14}}$}{\ifstrequal{#1}{A23B2C1E2G2I1}{$\tensor*{{\mathfrak{I}}}{_{26}}$}{\ifstrequal{#1}{A23B2C1E2G2I2}{$\tensor*{{\mathfrak{I}}}{_{27}}$}{\ifstrequal{#1}{A23B2C1E3}{$\tensor*{{\mathfrak{E}}}{_{16}}$}{\ifstrequal{#1}{A23B2C1E3F1}{$\tensor*{{\mathfrak{F}}}{_{16}}$}{\ifstrequal{#1}{A23B2C1E3F1H2}{$\tensor*{{\mathfrak{H}}}{_{25}}$}{\ifstrequal{#1}{A23B2C1E3F1H2J2}{$\tensor*{{\mathfrak{J}}}{_{13}}$}{\ifstrequal{#1}{A23B2C1E3F1H4}{$\tensor*{{\mathfrak{H}}}{_{24}}$}{\ifstrequal{#1}{A23B2C1E3G1}{$\tensor*{{\mathfrak{G}}}{_{24}}$}{\ifstrequal{#1}{A23B2C1E3G3}{$\tensor*{{\mathfrak{G}}}{_{25}}$}{\ifstrequal{#1}{A23B2C1E4}{$\tensor*{{\mathfrak{E}}}{_{17}}$}{\ifstrequal{#1}{A23B2C1E4G2}{$\tensor*{{\mathfrak{G}}}{_{26}}$}{\ifstrequal{#1}{A23B2C1E4G2H1}{$\tensor*{{\mathfrak{H}}}{_{35}}$}{\ifstrequal{#1}{A23B2C1E4G3}{$\tensor*{{\mathfrak{G}}}{_{27}}$}{\ifstrequal{#1}{A23B2D2}{$\tensor*{{\mathfrak{D}}}{_{5}}$}{\ifstrequal{#1}{A23B2D2F2}{$\tensor*{{\mathfrak{F}}}{_{13}}$}{\ifstrequal{#1}{A23B2D2F2H1}{$\tensor*{{\mathfrak{H}}}{_{21}}$}{\ifstrequal{#1}{A23B2D2F2H2}{$\tensor*{{\mathfrak{H}}}{_{22}}$}{\ifstrequal{#1}{A23B2D3}{$\tensor*{{\mathfrak{D}}}{_{6}}$}{\ifstrequal{#1}{A23B2D3F1}{$\tensor*{{\mathfrak{F}}}{_{14}}$}{\ifstrequal{#1}{A23B2D3F2}{$\tensor*{{\mathfrak{F}}}{_{15}}$}{\ifstrequal{#1}{A23B2D4}{$\tensor*{{\mathfrak{D}}}{_{7}}$}{\ifstrequal{#1}{A23B2D4F2}{$\tensor*{{\mathfrak{F}}}{_{17}}$}{\ifstrequal{#1}{A23B2D4F2G2}{$\tensor*{{\mathfrak{G}}}{_{31}}$}{\ifstrequal{#1}{A23B2D4F4}{$\tensor*{{\mathfrak{F}}}{_{18}}$}{\ifstrequal{#1}{A23B3}{$\tensor*{{\mathfrak{B}}}{_{3}}$}{\ifstrequal{#1}{A23B3D2}{$\tensor*{{\mathfrak{D}}}{_{8}}$}{\ifstrequal{#1}{A23B3D2F2}{$\tensor*{{\mathfrak{F}}}{_{19}}$}{\ifstrequal{#1}{A23B3D2F2H1}{$\tensor*{{\mathfrak{H}}}{_{27}}$}{\ifstrequal{#1}{A23B3D2F2H2}{$\tensor*{{\mathfrak{H}}}{_{28}}$}{\ifstrequal{#1}{A23B3D3}{$\tensor*{{\mathfrak{D}}}{_{9}}$}{\ifstrequal{#1}{A23B3D3F1}{$\tensor*{{\mathfrak{F}}}{_{20}}$}{\ifstrequal{#1}{A23B3D3F4}{$\tensor*{{\mathfrak{F}}}{_{21}}$}{\ifstrequal{#1}{A23B3D4}{$\tensor*{{\mathfrak{D}}}{_{10}}$}{\ifstrequal{#1}{A23B3D4F3}{$\tensor*{{\mathfrak{F}}}{_{22}}$}{\ifstrequal{#1}{A23B3D4F3G2}{$\tensor*{{\mathfrak{G}}}{_{32}}$}{\ifstrequal{#1}{A23B3D4F4}{$\tensor*{{\mathfrak{F}}}{_{23}}$}{\ifstrequal{#1}{A23B4}{$\tensor*{{\mathfrak{B}}}{_{4}}$}{\ifstrequal{#1}{A23B4D4}{$\tensor*{{\mathfrak{D}}}{_{11}}$}{\ifstrequal{#1}{A23B4D4F2}{$\tensor*{{\mathfrak{F}}}{_{24}}$}{\ifstrequal{#1}{A23B4D4F2G2}{$\tensor*{{\mathfrak{G}}}{_{33}}$}{\ifstrequal{#1}{A23B4D4F2H2}{$\tensor*{{\mathfrak{H}}}{_{29}}$}{\ifstrequal{#1}{A23B4D4F2H2J1}{$\tensor*{{\mathfrak{J}}}{_{15}}$}{\ifstrequal{#1}{A23B4D4F2H2J2}{$\tensor*{{\mathfrak{J}}}{_{16}}$}{\ifstrequal{#1}{A23B4D4F3}{$\tensor*{{\mathfrak{F}}}{_{25}}$}{\ifstrequal{#1}{A23B4D4F3H2}{$\tensor*{{\mathfrak{H}}}{_{30}}$}{\ifstrequal{#1}{A23B4D4F3H2J1}{$\tensor*{{\mathfrak{J}}}{_{17}}$}{\ifstrequal{#1}{A23B4D4F3H2J2}{$\tensor*{{\mathfrak{J}}}{_{18}}$}{\ifstrequal{#1}{A23B4D4F4}{$\tensor*{{\mathfrak{F}}}{_{26}}$}{\ifstrequal{#1}{A23B4D4F4H2}{$\tensor*{{\mathfrak{H}}}{_{31}}$}{\ifstrequal{#1}{A23B4D4F4H2J1}{$\tensor*{{\mathfrak{J}}}{_{19}}$}{\ifstrequal{#1}{A23B4D4F4H2J2}{$\tensor*{{\mathfrak{J}}}{_{20}}$}{\ifstrequal{#1}{A23B4D4G1}{$\tensor*{{\mathfrak{G}}}{_{3}}$}{\ifstrequal{#1}{A23B4D4G1I1}{$\tensor*{{\mathfrak{I}}}{_{6}}$}{\ifstrequal{#1}{A23B4D4G1I2}{$\tensor*{{\mathfrak{I}}}{_{7}}$}{\ifstrequal{#1}{A23B4D4G2}{$\tensor*{{\mathfrak{G}}}{_{4}}$}{\ifstrequal{#1}{A23B4D4G2I1}{$\tensor*{{\mathfrak{I}}}{_{8}}$}{\ifstrequal{#1}{A23B4D4G2I1J2}{$\tensor*{{\mathfrak{J}}}{_{24}}$}{\ifstrequal{#1}{A23B4D4G2I1J4}{$\tensor*{{\mathfrak{J}}}{_{25}}$}{\ifstrequal{#1}{A23B4D4G2I2}{$\tensor*{{\mathfrak{I}}}{_{9}}$}{\ifstrequal{#1}{A23B4D4G2I2J4}{$\tensor*{{\mathfrak{J}}}{_{26}}$}{\ifstrequal{#1}{A23B4D4G2I3}{$\tensor*{{\mathfrak{I}}}{_{10}}$}{\ifstrequal{#1}{A23B4E1}{$\tensor*{{\mathfrak{E}}}{_{3}}$}{\ifstrequal{#1}{A23B4E1G1}{$\tensor*{{\mathfrak{G}}}{_{9}}$}{\ifstrequal{#1}{A23B4E1G2}{$\tensor*{{\mathfrak{G}}}{_{10}}$}{\ifstrequal{#1}{A23B4E2}{$\tensor*{{\mathfrak{E}}}{_{4}}$}{\ifstrequal{#1}{A23B4E2G1}{$\tensor*{{\mathfrak{G}}}{_{11}}$}{\ifstrequal{#1}{A23B4E2G1H2}{$\tensor*{{\mathfrak{H}}}{_{32}}$}{\ifstrequal{#1}{A23B4E2G2}{$\tensor*{{\mathfrak{G}}}{_{12}}$}{\ifstrequal{#1}{A23C1}{$\tensor*{{\mathfrak{C}}}{_{1}}$}{\ifstrequal{#1}{A23C1E1}{$\tensor*{{\mathfrak{E}}}{_{5}}$}{\ifstrequal{#1}{A23C1E1G1}{$\tensor*{{\mathfrak{G}}}{_{13}}$}{\ifstrequal{#1}{A23C1E1G1I1}{$\tensor*{{\mathfrak{I}}}{_{11}}$}{\ifstrequal{#1}{A23C1E1G1I2}{$\tensor*{{\mathfrak{I}}}{_{12}}$}{\ifstrequal{#1}{A23C1E2}{$\tensor*{{\mathfrak{E}}}{_{6}}$}{\ifstrequal{#1}{A23C1E2G1}{$\tensor*{{\mathfrak{G}}}{_{14}}$}{\ifstrequal{#1}{A23C1E2G1I1}{$\tensor*{{\mathfrak{I}}}{_{13}}$}{\ifstrequal{#1}{A23C1E2G1I2}{$\tensor*{{\mathfrak{I}}}{_{14}}$}{\ifstrequal{#1}{A23C1E3}{$\tensor*{{\mathfrak{E}}}{_{7}}$}{\ifstrequal{#1}{A23C1E3G2}{$\tensor*{{\mathfrak{G}}}{_{15}}$}{\ifstrequal{#1}{A23C1E3G3}{$\tensor*{{\mathfrak{G}}}{_{16}}$}{\ifstrequal{#1}{A23C2}{$\tensor*{{\mathfrak{C}}}{_{2}}$}{\ifstrequal{#1}{A23C2E1}{$\tensor*{{\mathfrak{E}}}{_{8}}$}{\ifstrequal{#1}{A23C2E1F3}{$\tensor*{{\mathfrak{F}}}{_{27}}$}{\ifstrequal{#1}{A23C2E1G2}{$\tensor*{{\mathfrak{G}}}{_{17}}$}{\ifstrequal{#1}{A23C2E1G2I1}{$\tensor*{{\mathfrak{I}}}{_{15}}$}{\ifstrequal{#1}{A23C2E1G2I2}{$\tensor*{{\mathfrak{I}}}{_{16}}$}{\ifstrequal{#1}{A23C2E3}{$\tensor*{{\mathfrak{E}}}{_{9}}$}{\ifstrequal{#1}{A23C2E3G2}{$\tensor*{{\mathfrak{G}}}{_{18}}$}{\ifstrequal{#1}{A23C2E3G2I1}{$\tensor*{{\mathfrak{I}}}{_{17}}$}{\ifstrequal{#1}{A23C2E3G2I2}{$\tensor*{{\mathfrak{I}}}{_{18}}$}{\ifstrequal{#1}{A23C2F1}{$\tensor*{{\mathfrak{F}}}{_{1}}$}{\ifstrequal{#1}{A23C2F1H1}{$\tensor*{{\mathfrak{H}}}{_{1}}$}{\ifstrequal{#1}{A23C2F1H1I3}{$\tensor*{{\mathfrak{I}}}{_{20}}$}{\ifstrequal{#1}{A23C2F1H1I4}{$\tensor*{{\mathfrak{I}}}{_{23}}$}{\ifstrequal{#1}{A23C2F1H2}{$\tensor*{{\mathfrak{H}}}{_{2}}$}{\ifstrequal{#1}{A23C2F1H3}{$\tensor*{{\mathfrak{H}}}{_{3}}$}{\ifstrequal{#1}{A23C2F1H3I4}{$\tensor*{{\mathfrak{I}}}{_{22}}$}{\ifstrequal{#1}{A23C2F1H4}{$\tensor*{{\mathfrak{H}}}{_{4}}$}{\ifstrequal{#1}{A23C2F2}{$\tensor*{{\mathfrak{F}}}{_{2}}$}{\ifstrequal{#1}{A23C2F2H2}{$\tensor*{{\mathfrak{H}}}{_{5}}$}{\ifstrequal{#1}{A23C3}{$\tensor*{{\mathfrak{C}}}{_{3}}$}{\ifstrequal{#1}{A23C3E2}{$\tensor*{{\mathfrak{E}}}{_{10}}$}{\ifstrequal{#1}{A23C3E4}{$\tensor*{{\mathfrak{E}}}{_{11}}$}{\ifstrequal{#1}{A23C4}{$\tensor*{{\mathfrak{C}}}{_{4}}$}{\ifstrequal{#1}{A23C4E2}{$\tensor*{{\mathfrak{E}}}{_{12}}$}{\ifstrequal{#1}{A23C4E2F2}{$\tensor*{{\mathfrak{F}}}{_{28}}$}{\ifstrequal{#1}{A23C4E2G2}{$\tensor*{{\mathfrak{G}}}{_{19}}$}{\ifstrequal{#1}{A23C4E2G2I1}{$\tensor*{{\mathfrak{I}}}{_{19}}$}{\ifstrequal{#1}{A23C4E4}{$\tensor*{{\mathfrak{E}}}{_{13}}$}{\ifstrequal{#1}{A23C4E4G2}{$\tensor*{{\mathfrak{G}}}{_{20}}$}{\ifstrequal{#1}{A23C4E4G2I1}{$\tensor*{{\mathfrak{I}}}{_{21}}$}{\ifstrequal{#1}{A23C4F2}{$\tensor*{{\mathfrak{F}}}{_{3}}$}{\ifstrequal{#1}{A23C4F2H1}{$\tensor*{{\mathfrak{H}}}{_{6}}$}{\ifstrequal{#1}{A23Z1}{$\varnothing$}{\ifstrequal{#1}{S123}{$\aleph$}{\ifstrequal{#1}{S123B1}{$\tensor*{{\mathfrak{B}}}{_{1}}$}{\ifstrequal{#1}{S123B1C1}{$\tensor*{{\mathfrak{C}}}{_{5}}$}{\ifstrequal{#1}{S123B1C1D1}{$\tensor*{{\mathfrak{D}}}{_{1}}$}{\ifstrequal{#1}{S123B1C1D1F1}{$\tensor*{{\mathfrak{F}}}{_{1}}$}{\ifstrequal{#1}{S123B1C1D1F2}{$\tensor*{{\mathfrak{F}}}{_{2}}$}{\ifstrequal{#1}{S123B2}{$\tensor*{{\mathfrak{B}}}{_{2}}$}{\ifstrequal{#1}{S123C1}{$\tensor*{{\mathfrak{C}}}{_{1}}$}{\ifstrequal{#1}{S123C1D1}{$\tensor*{{\mathfrak{D}}}{_{2}}$}{\ifstrequal{#1}{S123C1D1F1}{$\tensor*{{\mathfrak{F}}}{_{3}}$}{\ifstrequal{#1}{S123C1D1F2}{$\tensor*{{\mathfrak{F}}}{_{4}}$}{\ifstrequal{#1}{S123C1D1F3}{$\tensor*{{\mathfrak{F}}}{_{5}}$}{\ifstrequal{#1}{S123C1D1F4}{$\tensor*{{\mathfrak{F}}}{_{6}}$}{\ifstrequal{#1}{S123C1E1}{$\tensor*{{\mathfrak{E}}}{_{1}}$}{\ifstrequal{#1}{S123C1E2}{$\tensor*{{\mathfrak{E}}}{_{2}}$}{\ifstrequal{#1}{S123C1E3}{$\tensor*{{\mathfrak{E}}}{_{3}}$}{\ifstrequal{#1}{S123C1E4}{$\tensor*{{\mathfrak{E}}}{_{4}}$}{\ifstrequal{#1}{S123C2}{$\tensor*{{\mathfrak{C}}}{_{2}}$}{\ifstrequal{#1}{S123C2E1}{$\tensor*{{\mathfrak{E}}}{_{5}}$}{\ifstrequal{#1}{S123C2E2}{$\tensor*{{\mathfrak{E}}}{_{6}}$}{\ifstrequal{#1}{S123C2E3}{$\tensor*{{\mathfrak{E}}}{_{7}}$}{\ifstrequal{#1}{S123C2E4}{$\tensor*{{\mathfrak{E}}}{_{8}}$}{\ifstrequal{#1}{S123C3}{$\tensor*{{\mathfrak{C}}}{_{3}}$}{\ifstrequal{#1}{S123C4}{$\tensor*{{\mathfrak{C}}}{_{4}}$}{\ifstrequal{#1}{S123Z1}{$\varnothing$}{$\tensor*{{\mathfrak{X}}}{_{0}}$}}}}}}}}}}}}}}}}}}}}}}}}}}}}}}}}}}}}}}}}}}}}}}}}}}}}}}}}}}}}}}}}}}}}}}}}}}}}}}}}}}}}}}}}}}}}}}}}}}}}}}}}}}}}}}}}}}}}}}}}}}}}}}}}}}}}}}}}}}}}}}}}}}}}}}}}}}}}}}}}}}}}}}}}}}}}}}}}}}}}}}}}}}}}}}}}}}}}}}}}}}}}}}}}}}}}}}}}}}}}}}}}}}}}}}}}}}
\newrobustcmd{\Lagrangian}[1]{%
	{\tensor*{\mathcal{L}}{_{\text{\NamedModel{#1}}}}}
}
\newrobustcmd{\UnknownMasslessParticle}{%
	{\big\{\tensor*{J}{^{P}_{\gamma}}\big\}}%
}
\newrobustcmd{\MasslessParticle}[2]{%
	{\tensor*{#1}{^{#2}_{\gamma}}}
}
\newrobustcmd{\MassiveParticle}[2]{%
	{\tensor*{#1}{^{#2}_{\text{m}}}}
}

\pgfkeys{
  /AnnotatedGraph/.is family, /AnnotatedGraph,
  A/.code n args={3}{ \node (labelA) at (#1, #3) {\huge$\mathfrak{A}_i$}; \draw[->,thick] (labelA.south) -- (#1, #2); },
  B/.code n args={3}{ \node (labelB) at (#1, #3) {\huge$\mathfrak{B}_i$}; \draw[->,thick] (labelB.south) -- (#1, #2); },
  C/.code n args={3}{ \node (labelC) at (#1, #3) {\huge$\mathfrak{C}_i$}; \draw[->,thick] (labelC.south) -- (#1, #2); },
  D/.code n args={3}{ \node (labelD) at (#1, #3) {\huge$\mathfrak{D}_i$}; \draw[->,thick] (labelD.south) -- (#1, #2); },
  E/.code n args={3}{ \node (labelE) at (#1, #3) {\huge$\mathfrak{E}_i$}; \draw[->,thick] (labelE.south) -- (#1, #2); },
  F/.code n args={3}{ \node (labelF) at (#1, #3) {\huge$\mathfrak{F}_i$}; \draw[->,thick] (labelF.south) -- (#1, #2); },
  G/.code n args={3}{ \node (labelG) at (#1, #3) {\huge$\mathfrak{G}_i$}; \draw[->,thick] (labelG.south) -- (#1, #2); },
  H/.code n args={3}{ \node (labelH) at (#1, #3) {\huge$\mathfrak{H}_i$}; \draw[->,thick] (labelH.south) -- (#1, #2); },
  I/.code n args={3}{ \node (labelI) at (#1, #3) {\huge$\mathfrak{I}_i$}; \draw[->,thick] (labelI.south) -- (#1, #2); },
  J/.code n args={3}{ \node (labelJ) at (#1, #3) {\huge$\mathfrak{J}_i$}; \draw[->,thick] (labelJ.south) -- (#1, #2); },
  K/.code n args={3}{ \node (labelK) at (#1, #3) {\huge$\mathfrak{K}_i$}; \draw[->,thick] (labelK.south) -- (#1, #2); },
  L/.code n args={3}{ \node (labelL) at (#1, #3) {\huge$\mathfrak{L}_i$}; \draw[->,thick] (labelL.south) -- (#1, #2); },
  M/.code n args={3}{ \node (labelM) at (#1, #3) {\huge$\mathfrak{M}_i$}; \draw[->,thick] (labelM.south) -- (#1, #2); },
  N/.code n args={3}{ \node (labelN) at (#1, #3) {\huge$\mathfrak{N}_i$}; \draw[->,thick] (labelN.south) -- (#1, #2); },
}

\newrobustcmd{\AnnotatedGraph}[2][]{%
  \begin{tikzpicture}
    \node[anchor=south west, inner sep=0] (image) at (0,0) {\includegraphics[width=\linewidth]{#2}};

    \begin{pgfonlayer}{foreground}
        \begin{scope}[
            x={($(image.south east)-(image.south west)$)},
            y={($(image.north west)-(image.south west)$)}
        ]
          \pgfkeys{/AnnotatedGraph, #1}
        \end{scope}
    \end{pgfonlayer}
  \end{tikzpicture}%
}

\usepackage{hyperref}
\hypersetup{%
colorlinks = true,%
linkcolor = Red,%
citecolor = Red,%
filecolor = Red,%
urlcolor = Red%
}%
\usepackage[capitalize]{cleveref}
\AddToHook{cmd/appendix/before}{\crefalias{section}{appendix}}
\makeatletter
\renewcommand{\paragraph}{%
\@startsection{paragraph}{4}%
{\z@}{1.21ex \@plus 1ex \@minus .2ex}{0.9em}%
{\normalfont\normalsize\bfseries}%
}
\usepackage{titlesec}
\newrobustcmd{\pea}[1]{\emph{#1}\textbf{.\ \ \ ---}}
\titleformat{\paragraph}[runin]{\normalfont\normalsize\bfseries}{\emph\theparagraph}{1em}{\pea}
\titleformat{\section}[block]{\normalfont\bfseries\centering}{\MakeUppercase\thesection}{1em}{\MakeUppercase}
\makeatother

\makeatletter \def\switch@array{}\makeatother

\allowdisplaybreaks

\begin{document}

\title{Infrared foundations for quantum geometry I:\\
Catalogue of totally symmetric rank-three field theories}

\author{Will Barker}
\email{barker@fzu.cz}
\affiliation{Central European Institute for Cosmology and Fundamental Physics, Institute of Physics of the Czech Academy of Sciences, Na Slovance 1999/2, 182 00 Prague 8, Czechia}

\author{Carlo Marzo}
\email{carlo.marzo@kbfi.ee}
\affiliation{Laboratory for High Energy and Computational Physics, NICPB, R\"{a}vala 10, Tallinn 10143, Estonia}

\author{Alessandro Santoni}
\email{asantoni@uc.cl}
\affiliation{Institut f\"ur Theoretische Physik, Technische Universit\"at Wien, Wiedner Hauptstrasse 8--10, A-1040 Vienna, Austria}
\affiliation{Facultad de F\'isica, Pontificia Universidad Cat\'olica de Chile, Vicu\~{n}a Mackenna 4860, Santiago, Chile}

\begin{abstract}
	We systematically obtain all linear models which propagate a totally symmetric rank-three field without parity violation on a flat background. Each such model is defined exclusively by its gauge symmetry, a necessary property of effective field theories in the infrared limit. By comparison, models obtained by other means (tuning couplings or cherry-picking operators) may be unstable against radiative corrections. For each model, we compute the spectrum of massless and massive particles, and the no-ghost-no-tachyon constraints on the couplings. We conclude that foundational models exist which can propagate one massless particle of spin one or spin three in isolation, or both particles simultaneously, generalising the model of Campoleoni and Francia. Our algorithm for detecting symmetric models is grounded in particle physics methods, being based directly on the Wigner decomposition of the field. Compared to our recent analysis of the totally symmetric rank-three field (whose results we confirm and extend) our new algorithm does not require an ansatz for the symmetry transformation, and is not restricted to so-called `free' symmetries.
\end{abstract}

\maketitle

\tableofcontents

\begin{figure}[h]
\includegraphics[width=\linewidth]{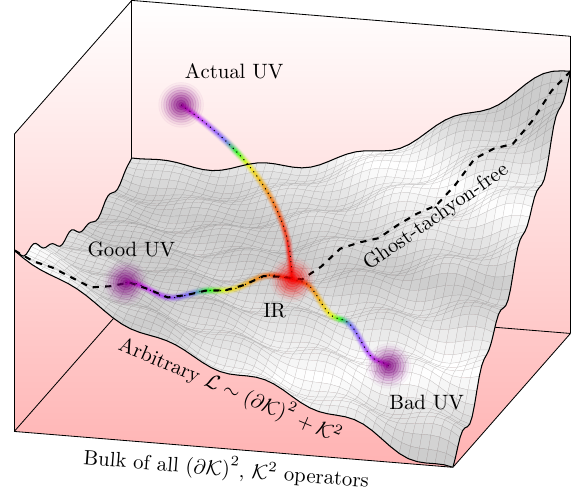}
\caption{\label{Propaganda} Consider a low-energy theory defined by an arbitrary sub-set of cherry-picked operators. Radiative corrections will de-tune the no-ghost-no-tachyon constraints on this theory or, worse, require the addition of excluded operators so as to absorb un-cancelled ultraviolet divergences. These effects are prevented by tuning couplings to the minimum extent needed to produce some gauge symmetry.}
\end{figure}

\begin{table*}[htbp]
	\includegraphics[width=\linewidth]{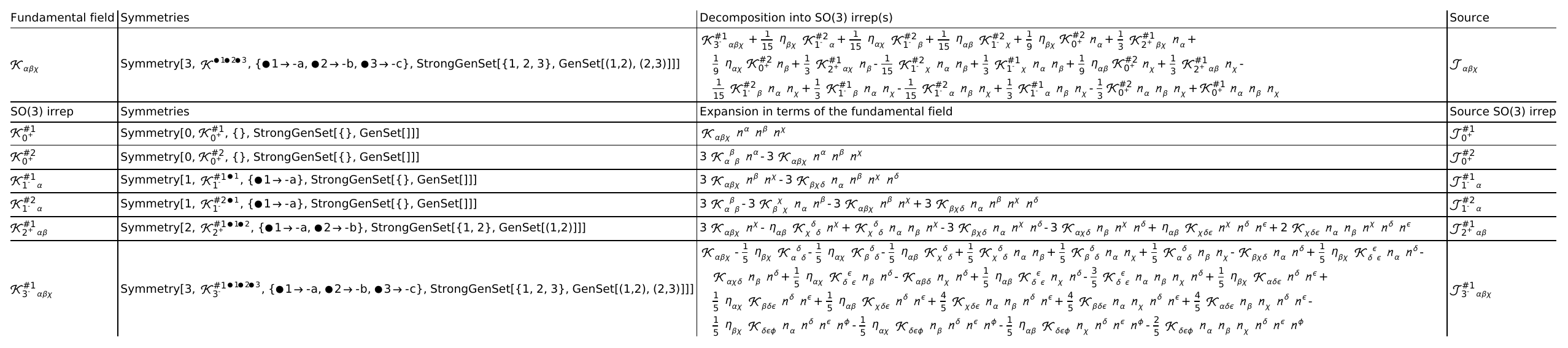}
	\caption{Output generated by \PSALTer{}. The field kinematics of the totally symmetric field defined in~\cref{eq:DisS123}. These definitions are used in~\cref{ParticleSpectrographS123,ParticleSpectrographS123C1E1,ParticleSpectrographS123C1E2,ParticleSpectrographS123C2E2,ParticleSpectrographS123B1C1D1F2,ParticleSpectrographS123C1D1F2}. The unit-timelike vector is~$\tensor{n}{_{\mu}}\equiv\tensor{k}{_{\mu}}/\sqrt{\tensor{k}{^\nu}\tensor{k}{_\nu}}$, where~$\tensor{k}{_\mu}$ is the massive particle four-momentum.}
	\label{FieldKinematicsS123Field}
\end{table*}

\section{Introduction}\label{Sec:Introduction}

\paragraph*{Effective field theory} As explained in~\cite{Barker:2025xzd} and illustrated in~\cref{Propaganda}, attempts to perturbatively bootstrap a viable theory of some field~$\Dis{}$ are only expected to succeed if \emph{all} operators are included which are consistent with the symmetries of the theory. The bootstrapping procedure assumes a simultaneous expansion in powers of the canonically normalized and perturbative field~$\Dis{}$ and its derivatives, i.e. the low-energy expansion
\begin{align}\label{EFTLag}
\mathcal{L}=\mathcal{L}_{n\leq4}+\sum_{n=5}^{\infty}\sum_i\frac{\MAGG{n}{_i}\OperatorO{i}{n}}{\Lambda^{n-4}} \,,
\end{align}
where the~$\big\{\MAGG{n}{_i}\big\}$ are dimensionless couplings for which~$i$ labels multiple operators with the same mass dimension~$n$, and the constant~$\Lambda$ is a cutoff scale. When canonical normalization assumes~$\Dis{}$ to have mass-dimension one, and no auxiliary fields are needed to eliminate higher-derivatives, the starting-point of the bootstrap is the collection of relevant and marginal operators which are quadratic in~$\Dis{}$ and have up to two derivatives, i.e. the Lagrangian~$\mathcal{L}_{n\leq4}$ in~\cref{EFTLag}. If~$\Dis{}$ is a high-rank tensor field and no extra symmetries are specified, the general ansatz for~$\mathcal{L}_{n\leq4}$ is unlikely to be unitary. What this observation tells us is that~$\Dis{}$ alone is not a foundation for a physical theory. Good foundations may be found by additionally specifying a gauge symmetry, if the symmetric restriction of~$\mathcal{L}_{n\leq4}$ uniquely leads to a unitary model. What this observation does \emph{not} tell us, is that we are simply allowed to tune the couplings in~$\mathcal{L}_{n\leq4}$ so as to arrive at a unitary spectrum of propagating degrees of freedom (d.o.f). The latter `cherry-picking' approach can only be expected to work when it happens to produce a model that can also be reached by the former `symmetrization' approach. Otherwise, there is no generic expectation that the tuned couplings will remain stable under radiative corrections. Rather than falling into the generally renormalizable or non-renormalizable structure of~\cref{EFTLag} -- which is physical and predictive -- arbitrarily tuned models are in danger of being \emph{not-even-non-renormalizable}, i.e. not being physical in any sense~\cite{Georgi:1993mps,Burgess:2007pt,Bijnens:2006zp,Manohar:1996cq,Kaplan:1995uv,Pich:1995bw,Ecker:1994gg,Dobado:1989ax,Dobado:1989gr,Weinberg:1978kz,Coleman:1969sm,Callan:1969sn}.

\paragraph*{In this letter} We consider as~$\Dis{}$ a totally symmetric rank-three field of the form
\begin{equation}\label{eq:DisS123}
	\Dis{_{\alpha\beta\chi}}\equiv\Dis{_{(\alpha\beta\chi)}} \, .
\end{equation}
Such fields may have various motivations and interpretations, though these are not needed and are not necessarily productive. For example, it has been proposed that the gravitational connection may contain an independent, dynamical rank-three tensor part~\cite{Baldazzi:2021kaf,Martini:2023apm,Sauro:2022chz,Sauro:2024ujx,Melichev:2023lwj,Melichev:2025hcg}, for which the totally symmetric case in~\cref{eq:DisS123} was considered in~\cite{Iosifidis:2023wbx}. In any case, up to boundary terms the most general theory (denoted~\NamedModel{S123} to distinguish it from its special cases) built from the relevant and marginal quadratic operators of the field defined in~\cref{eq:DisS123} has Lagrangian
\begin{align}
	\Lagrangian{S123} & \equiv
	\MAGG{2}{_1}\Lambda^2\Dis{_{\alpha\beta\chi}}\Dis{^{\alpha\beta\chi}} 
	+\MAGG{2}{_2}\Lambda^2\Dis{^{\alpha}_{\alpha}^{\beta}}\Dis{_{\beta}^{\chi}_{\chi}}
	\nonumber\\& \hspace{5pt}
	+\MAGG{4}{_1}\PD{_{\beta}}\Dis{_{\chi}^{\delta}_{\delta}}
		\PD{^{\chi}}\Dis{^{\alpha}_{\alpha}^{\beta}}
	+\MAGG{4}{_2}\PD{_{\chi}}\Dis{_{\beta}^{\delta}_{\delta}}
		\PD{^{\chi}}\Dis{^{\alpha}_{\alpha}^{\beta}}
	\nonumber\\& \hspace{5pt}
	+\MAGG{4}{_3}\PD{_{\alpha}}\Dis{^{\alpha\beta\chi}}
		\PD{_{\delta}}\Dis{_{\beta\chi}^{\delta}}
	+\MAGG{4}{_4}\PD{^{\chi}}\Dis{^{\alpha}_{\alpha}^{\beta}}\PD{_{\delta}}\Dis{_{\beta\chi}^{\delta}}
	\nonumber\\& \hspace{5pt}
	+\MAGG{4}{_7}\PD{_{\delta}}\Dis{_{\alpha\beta\chi}}
		\PD{^{\delta}}\Dis{^{\alpha\beta\chi}}
	\, , \label{RootTheoryS123}
\end{align}
requiring the canonical dimension of~$\Dis{_{\alpha\beta\chi}}$ to be unity, and neglect parity-violating operators. This latter restriction is somewhat arbitrary, since parity is known already to be maximally violated in the weak sector, and it may be relaxed in an extended analysis without modifying the methods of this letter~\cite{Barker:2025qmw}. The case of a fully symmetric tensor as in~\cref{eq:DisS123} will demonstrate how consistency requirements significantly narrow down the space of viable quadratic actions. This feature is particularly advantageous, echoing the stringent consistency requirements that uniquely single out, for lower-rank cases, the theories of Maxwell and Einstein. The outcome for the current higher-rank case is both transparent and consequential: gauge symmetries, essential for excluding pathological states, strictly permit only spin-three and spin-one propagation. The letter is organised as follows. In~\cref{Sec:Methods} we describe the algorithm used to obtain physically motivated special cases of~\cref{RootTheoryS123}. In~\cref{Sec:Concl} we present the results of the algorithm and discuss their physical significance. The full catalogue of models is given in~\cref{Sec:Parameters}.

\begin{figure*}[htbp]
	\includegraphics[width=\linewidth]{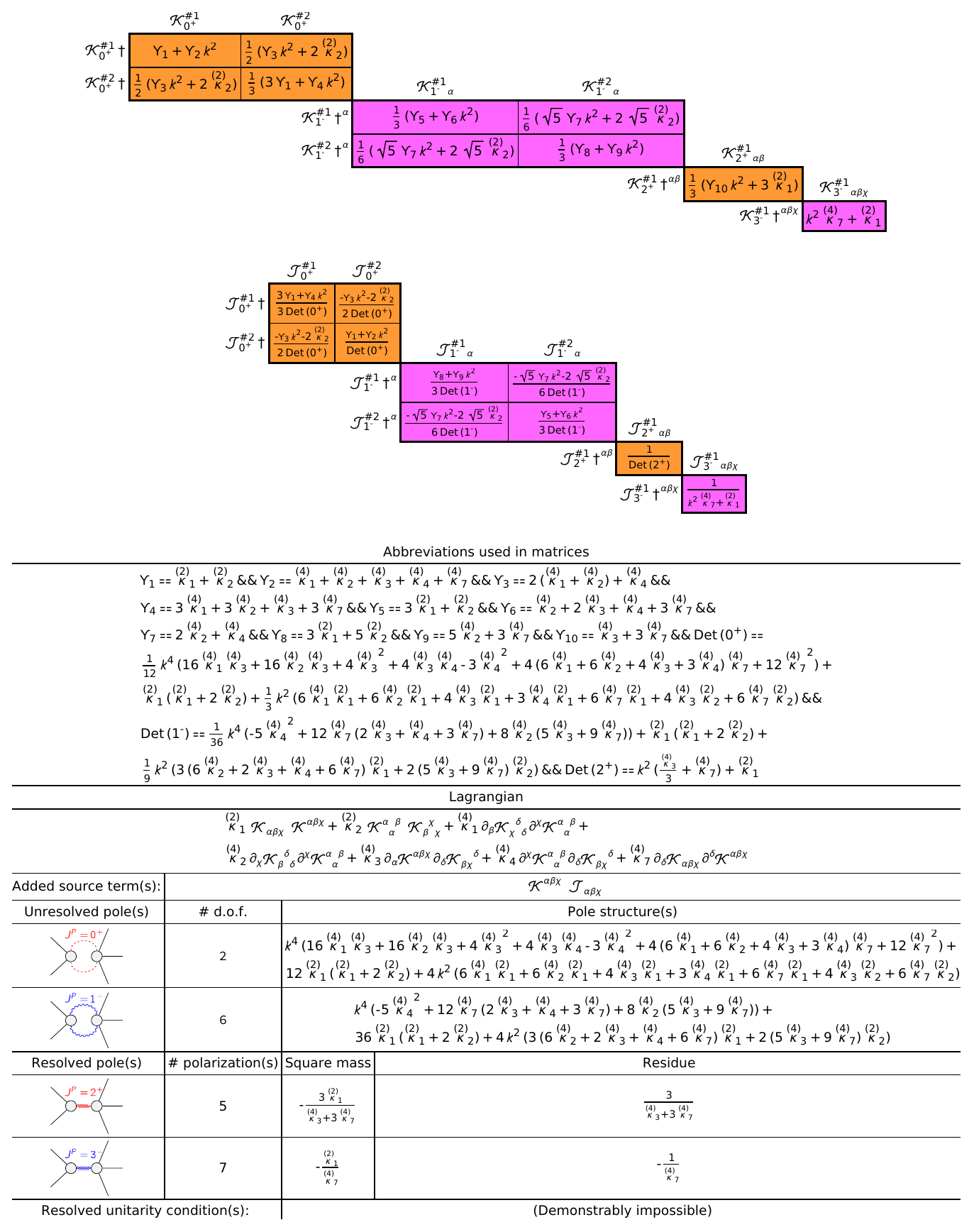}
	\caption{Output generated by \PSALTer{}. The spectrograph of~\NamedModel{S123}, as defined in~\cref{RootTheoryS123}. All notation is defined in~\cref{FieldKinematicsS123Field}. Note that when presenting spectrographs, we absorb the factor~$\Lambda^2$ into the~$\big\{\MAGG{2}{_i}\big\}$. The upper matrices denote the blocks~$\mathsf{O}_{J^P}$ in~\cref{BlockDiagonal}, and the lower matrices denote the pseudoinverses~$\mathsf{O}^+_{J^P}$. The model is not unitary.}
\label{ParticleSpectrographS123}
\end{figure*}

\paragraph*{Conventions} We adopt natural units in which~$\hbar\equiv c\equiv 1$, and the particle physics signature~$(+,-,-,-)$.

\section{Theoretical development}\label{Sec:Methods}

\paragraph*{Particle spectroscopy} As we uncover symmetric special cases of~\cref{RootTheoryS123}, it will be important to compute their tree-level spectra. For this task, we apply the \textit{PSALTer} software~\cite{Barker:2024juc,Barker:2025qmw}. To study the propagator associated with some Lagrangian~$\mathcal{L}_{n\leq4}$ depending only on~$\Dis{_{\alpha\beta\chi}}$ and its derivatives, and parameterised by the couplings~$\big\{\MAGG{n}{_i}\big\}$ and cutoff~$\Lambda$, a formal collection of test sources~$\Cur{^{\alpha\beta\chi}}$ is introduced to the action~$\mathcal{S}\equiv\int\mathrm{d}^4x\big[\mathcal{L}_{n\leq4}-\Dis{_{\alpha\beta\chi}}\Cur{^{\alpha\beta\chi}}\big]$. In programmatical terms we represent the independent d.o.f in~$\Dis{_{\alpha\beta\chi}}$ and~$\Cur{^{\alpha\beta\chi}}$ with the column vectors~$\mathsf{J}$ and the conjugate field~$\mathsf{K}$. In position space, this action is then expressed schematically as
\begin{equation}\label{GenLag}
	\mathcal{S}=\int\mathrm{d}^4x\ \mathsf{K}^{\text{T}}(x)\cdot\Big[\mathsf{O}(\partial)\cdot\mathsf{K}(x)-\mathsf{J}(x)\Big].
\end{equation}
The operator~$\mathsf{O}(\partial)$ is the wave operator: a tensor polynomial of (up to) quadratic order in~$\PD{_\mu}$ and parameterised by the~$\big\{\MAGG{n}{_i}\big\}$ and~$\Lambda$. The saturated propagator associated with~\cref{GenLag} is obtained in momentum space~$\PD{_\mu}\mapsto -i\tensor{k}{_\mu}$ by `inverting' the wave operator and sandwiching it between the sources according to
\begin{equation}\label{Propagator}
	\Pi(k)\equiv\mathsf{J}^\dagger(k)\cdot\mathsf{O}^{-1}(k)\cdot\mathsf{J}(k).
\end{equation}
By construction, the special cases of~\cref{RootTheoryS123} which we consider will all have extra gauge symmetries: when such symmetries are present the `inverse' in~\cref{Propagator} does not actually exist due to degeneracies of the wave operator. As we will see presently, precisely those symmetries that are responsible for the degeneracies also result in constraints on the physical (on-shell) parts of the source currents. Accordingly, the result of the `sandwiching' in~\cref{Propagator} is that the infinities always cancel against the zeros. According to the usual principles of polology, the positions of the poles in~$\Pi(k)$ determine the masses of physical states, with real masses needed for non-tachyonic particles. The residues of~$\Pi(k)$ on each pole must be positive definite in order to avoid a ghost. For massive states, adopting a spin-parity ($J^P$) basis is beneficial for casting the wave operator inversion as a linear algebra problem. In this basis, the wave operator decomposes into irreducible subspaces
\begin{equation}\label{BlockDiagonal}
	\mathsf{O}(k)\equiv\bigoplus_{J,P}\tensor{\mathsf{O}}{_{J^P}}(k),
\end{equation}
and the spectroscopy can be performed in each subspace separately. At the point of definition, \PSALTer{} automatically determines the~$J^P$ sectors associated with~$\Dis{_{\alpha\beta\chi}}$, as illustrated in~\cref{FieldKinematicsS123Field}. Equipped with this basis, the spectrograph of the unconstrained~\NamedModel{S123} theory in~\cref{RootTheoryS123} is shown in~\cref{ParticleSpectrographS123}. In the unconstrained case, it turns out that \emph{all} the poles are massive. Each of the~$0^+$ and~$1^-$ sectors of~\NamedModel{S123} propagates a \emph{pair} of particles. The expressions for the squares of the~$0^+$ and~$1^-$ masses are not rational functions of the~$\big\{\MAGG{n}{_i}\big\}$, since they are obtained by solving a quadratic in~$k\equiv\sqrt{\tensor{k}{^\nu}\tensor{k}{_\nu}}$. To avoid long calculations involving square roots, \PSALTer{} sets these poles aside as `unresolved'. The `resolved' poles, however, are sufficient to understand the sick nature of the~\NamedModel{S123} theory. There are isolated poles in the~$2^+$ and~$3^-$ sectors of the theory whose mutual no-ghost-no-tachyon constraints are contradictory. As we have argued in~\cref{Sec:Introduction}, the only path forward in such cases is to introduce gauge symmetries. This process will naturally eliminate masses, as is the case with the Fierz--Pauli and Proca--Maxwell models, and so we will have to contend with the massless polology.

\begin{figure*}[htbp]
\includegraphics[width=\linewidth]{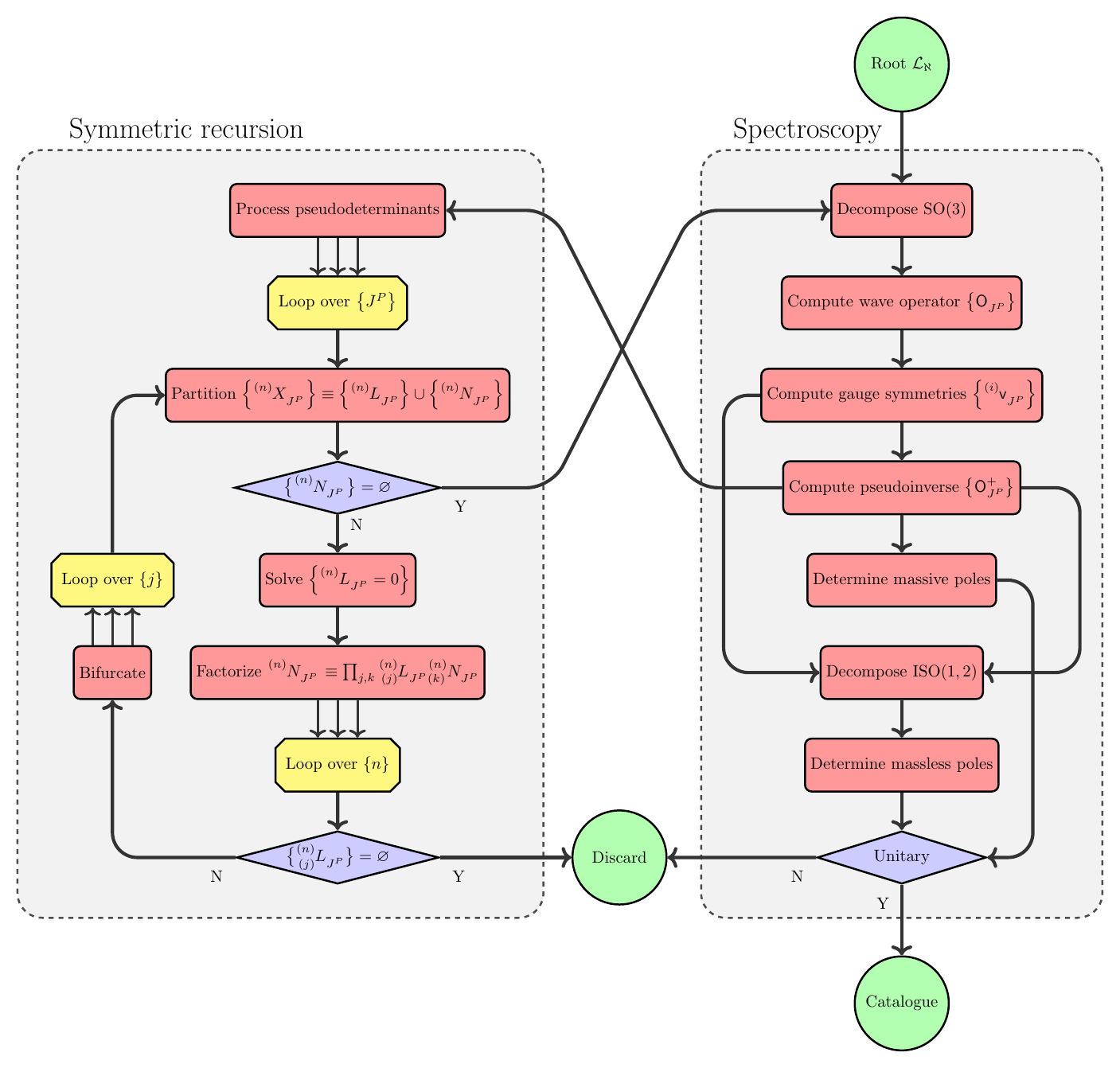}
	\caption{\label{Algorithm} The algorithm takes the general theory~$\mathcal{L}_{\aleph}$ as input, as defined in~\cref{RootTheoryS123}. The \PSALTer{} software is used to perform the spectroscopy of the model (right), and this is integrated with a new routine for obtaining special cases with enhanced gauge symmetries (left). The existing implementation in \PSALTer{} simultaneously assesses the unitarity of each model as a final output, and performs the spin-parity decomposition of the wave operator needed by the new routine as raw input for inducing new symmetries. The final science product is an exhaustive catalogue of all physically-motivated special cases of the general theory (see~\cref{GraphRepresentationS123,AllModelsS123}).}
\end{figure*}

\paragraph*{Helicity extension} Whilst massive states can be fully understood using the~$J^P$ basis, in the case of massless states spin~$J$ is replaced by helicity~$h$. The existing implementation in \PSALTer{} proceeds component-wise with the massless computation in the fiducial frame~$\left[\tensor{k}{^\mu}\right] = \left(\omega, 0, 0, \omega\right)$. This approach is guaranteed to determine the number of massless polarisations, and their unitarity conditions, but it does not inform about the~$J$ or~$h$ quantum numbers. For the first time in this letter, we extend beyond the theory developed in~\cite{Barker:2024juc,Barker:2025qmw} by interpreting the component-wise results from \PSALTer{} in terms of the physical helicity decomposition, i.e. in terms of states of definite angular momentum. As a basic example, we consider the case of a conjugate source~$\Cur{^\mu}$ to some vector field~$\Dis{_\mu}$. In the fiducial light-like frame the source has components~$\left[\Cur{^\mu}(k)\right]=(\Cur{^0}(k),\Cur{^1}(k), \Cur{^2}(k), \Cur{^3}(k))$, and the well-known combinations~$\hel{\pm 1}(k) \equiv \Cur{^1}(k) \pm i \Cur{^2}(k)$ clearly possess definite helicities of value~$\pm 1$. Extending this example to the present case, straightforward decomposition rules for angular momentum in the reducible tensorial representation~$\Cur{^{\alpha \beta \gamma}}(k)$ yield the analogous formulae 
\begin{subequations}
\begin{align} 
	\hel{\pm 3} &\equiv \Cur{^{222}} -3 \Cur{^{112}} \pm i( 3 \Cur{^{122}} - \Cur{^{111}}), \label{Rank3HelFirst} \\ 
	\hel[1]{\pm 2} &\equiv \Cur{^{223}} - \Cur{^{113}}  \pm 2 i \Cur{^{123}},  \\
	\hel[2]{\pm 2} &\equiv \Cur{^{022}} - \Cur{^{011}}  \pm 2 i \Cur{^{012}},  \\
	\hel[1]{\pm 1} &\equiv \Cur{^{233}} \pm i \Cur{^{133}},   \\
	\hel[2]{\pm 1} &\equiv \Cur{^{112}} + \Cur{^{222}} \pm i(\Cur{^{111}} + \Cur{^{122}}), \\
	\hel[3]{\pm 1} &\equiv \Cur{^{023}} \pm i \Cur{^{013}},  \\
	\hel[4]{\pm 1} &\equiv \Cur{^{002}} \pm i \Cur{^{001}},  \\
	\hel[1]{0} &\equiv \Cur{^{333}},  \\
	\hel[2]{0} &\equiv \Cur{^{113}} + \Cur{^{223}},  \\
	\hel[3]{0} &\equiv \Cur{^{033}},  \\
	\hel[4]{0} &\equiv \Cur{^{011}} + \Cur{^{022}},  \\
	\hel[5]{0} &\equiv \Cur{^{003}},  \\
	\hel[6]{0} &\equiv \Cur{^{000}}. \label{Rank3HelLast}
\end{align}
\end{subequations}
For the 20 independent components of the totally symmetric source~$\Cur{^{\alpha \beta \chi}}(k)$, the invariant combinations corresponding to definite helicity values are denoted by the symbols~$\hel[d]{h}$. The subscript~$h$ indicates the helicity value, while the superscript distinguishes between multiple degenerate eigenvectors, where present. In our case, these formulae must further account for reductions in d.o.f imposed by the gauge constraints specific to each model we find. Information about gauge constraints is already managed internally by the component-wise procedure in \PSALTer{}, and this allows us to construct simultaneously helicity- and gauge-invariant combinations of source components. Applying the reverse transformation from \crefrange{Rank3HelFirst}{Rank3HelLast} to the saturated propagator~$\Pi(k)$ yields a simplified expression, conclusively resolving the identification of the massless propagating states. This method is not yet systematized in \PSALTer{}, and so the massless spectrographs in~\cref{ParticleSpectrographS123C1E1,ParticleSpectrographS123C1E2,ParticleSpectrographS123C2E2,ParticleSpectrographS123B1C1D1F2,ParticleSpectrographS123C1D1F2} do not refer to the helicity. Rather, we apply the method manually to confirm our claims in~\cref{Sec:Concl} about the $J$-sector of origin for each massless mode.

\paragraph*{Symmetric recursion} For this letter we augment the method of particle spectroscopy with a recursive search algorithm. Working from~\cref{GenLag}, the Fourier space shell of classical field equations is
\begin{equation}\label{FieldEquations}
	\mathsf{O}(k)\cdot\mathsf{K}(k)=\frac{1}{2}\mathsf{J}(k).
\end{equation}
If~$\mathsf{K}(k)$ is a solution to~\cref{FieldEquations}, then so is~$\mathsf{K}(k)+\delta\mathsf{K}(k)$, where~$\delta\mathsf{K}\equiv\sum_ic_i(k)\GenNullVector{i}(k)$ is spanned by the collection of all null right eigenvectors~$\mathsf{O}(k)\cdot\GenNullVector{i}(k)\equiv0$. Such shifts evidently describe the gauge symmetries of the theory, for which~$c_i(x)$ are the independent local generators. By physicality, the wave operator~$\mathsf{O}(k)$ is assumed to be Hermitian, so that the right eigenvectors~$\GenNullVector{i}(k)$ are also left eigenvectors~$\GenNullVector{i}^\dagger(k)\cdot\mathsf{O}(k)\equiv 0$. Left-multiplying~\cref{FieldEquations} by~$\GenNullVector{i}^\dagger(k)$ gives~$\GenNullVector{i}^\dagger(k)\cdot\mathsf{J}(k)=0$, which defines the on-shell constraints on the sources. It is the existence of the collection~$\left\{\GenNullVector{i}(k)\right\}$ and, accordingly, the existence of one or more zero eigenvalues of~$\mathsf{O}(k)$ that makes the inverse in~\cref{Propagator} ill-defined; it is the source constraints which prevent this from being a problem. Accordingly, it follows that the poles in the saturated propagator may be identified by the pseudodeterminant~$\Pi(k)\propto\det\tilde{\mathsf{O}}^{-1}(k)$ where the invertible matrix~$\tilde{\mathsf{O}}$ is defined as
\begin{equation}\label{PseudoDeterminant}
	\tilde{\mathsf{O}}(k)\equiv\mathsf{O}(k)+\sum_i\GenNullVector{i}(k)\cdot\GenNullVector{i}^\dagger(k).
\end{equation}
In general, we have the polynomial structure~$\det\tilde{\mathsf{O}}(k)=\sum_n \GenXExpr{n}k^n$ for~$n\geq 0$, and some collection~$\left\{\GenXExpr{n}\right\}$ of coefficients~$\GenXExpr{n}$ which are themselves polynomials in the couplings. Starting from the most general possible ansatz of quadratic operators, every solution to the simultaneous system~$\left\{\GenXExpr{n}=0\right\}$ defines a distinct theory whose only motivating principle is the emergence of one or more new gauge symmetry. There is no reason why the~$\left\{\GenXExpr{n}\right\}$ should be linear in the couplings. In those non-linear cases which factorise, the solutions to~$\left\{\GenXExpr{n}=0\right\}$ will bifurcate into multiple branches, each of which corresponds to a distinct theory. To study all the branches, it is prudent to work with any linear equations for as long as possible, since all equations in the system must anyway be satisfied in the end. One may partition the various expressions into linear and non-linear parts 
\begin{align}
	\left\{\GenXExpr{n}\right\} & \equiv \left\{\GenLExpr{n}\right\}\cup\left\{\GenNExpr{n}\right\}.
\end{align}
The system~$\big\{\GenLExpr{n}=0\big\}$ can be uniquely solved, and new bifurcation points will become apparent if the unique solution allows any of the~$\big\{\GenNExpr{n}\big\}$ to factorize algebraically. If any of the new factors are linear, this process repeats. For a given~$\mathsf{O}(k)$ one is ideally left with a collection of models defined \emph{entirely} by linear constraints on the couplings: these constraints are not arbitrary, because they are independent of coupling redefinitions. The algorithm may also leave residual cases which resist factorisation, and which are defined by inherently non-linear constraints: each non-linear constraint can be `sliced' into two or more linear constraints, but the infinite choice of slicings makes this arbitrary. Fortunately, the inherently non-linear models are relatively few in the system following from~\cref{RootTheoryS123}. In practice, as with the spectroscopy itself, the symmetric recursion algorithm is easiest to implement in the~$J^P$ basis. With reference to~\cref{BlockDiagonal}, the search proceeds independently in each subspace~$\tensor{\mathsf{O}}{_{J^P}}(k)$, with null vectors~$\big\{\NullVector{J}{P}{i}(k)\big\}$ giving rise to~$\big\{\XExpr{J}{P}{n}\big\}$,~$\big\{\LExpr{J}{P}{n}\big\}$ and~$\big\{\NExpr{J}{P}{n}\big\}$. We illustrate this algorithm in~\cref{Algorithm}.

\begin{figure}[htbp]
	\AnnotatedGraph[%
		B = {0.305}{0.7}{1.07},%
		C = {0.435}{0.8}{1.07},%
		D = {0.565}{1.}{1.07},%
		E = {0.695}{0.9}{1.07},%
		F = {0.825}{0.81}{1.07},%
	]{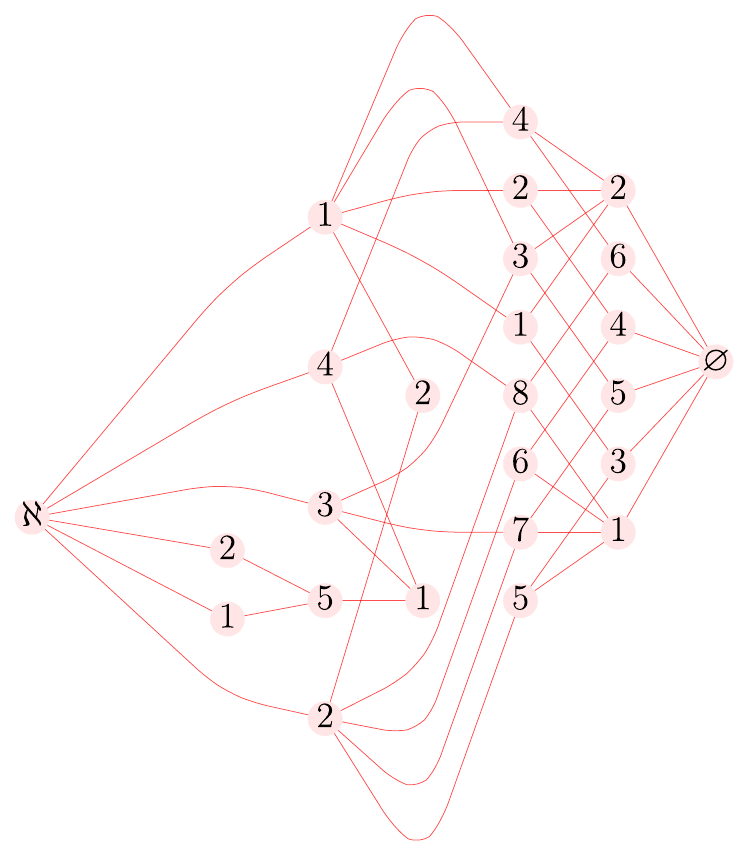}
	\caption{\label{GraphRepresentationS123} Catalogue of ways to propagate a totally symmetric rank-three field. Each node in the graph represents a distinct theory, defined by imposing constraints (from left to right) on the unconstrained root theory~\NamedModel{S123} defined in~\cref{RootTheoryS123}. The final node~\NamedModel{S123Z1} is the trivial theory~$\Lagrangian{S123Z1}\equiv 0$. Constraints that connect a given daughter theory to its nearest parent are represented by edges. All these models are listed in~\cref{AllModelsS123}. Among them, the cases~\NamedModel{S123C1E1},~\NamedModel{S123C1E2},~\NamedModel{S123C2E2},~\NamedModel{S123B1C1D1F2}, and~\NamedModel{S123C1D1F2} are unitary, see~\cref{ParticleSpectrographS123C1E1,ParticleSpectrographS123C1E2,ParticleSpectrographS123C2E2,ParticleSpectrographS123B1C1D1F2,ParticleSpectrographS123C1D1F2}.}
\end{figure}

\section{Results and discussion}\label{Sec:Concl}

\begin{figure*}[htbp]
	\includegraphics[width=\linewidth]{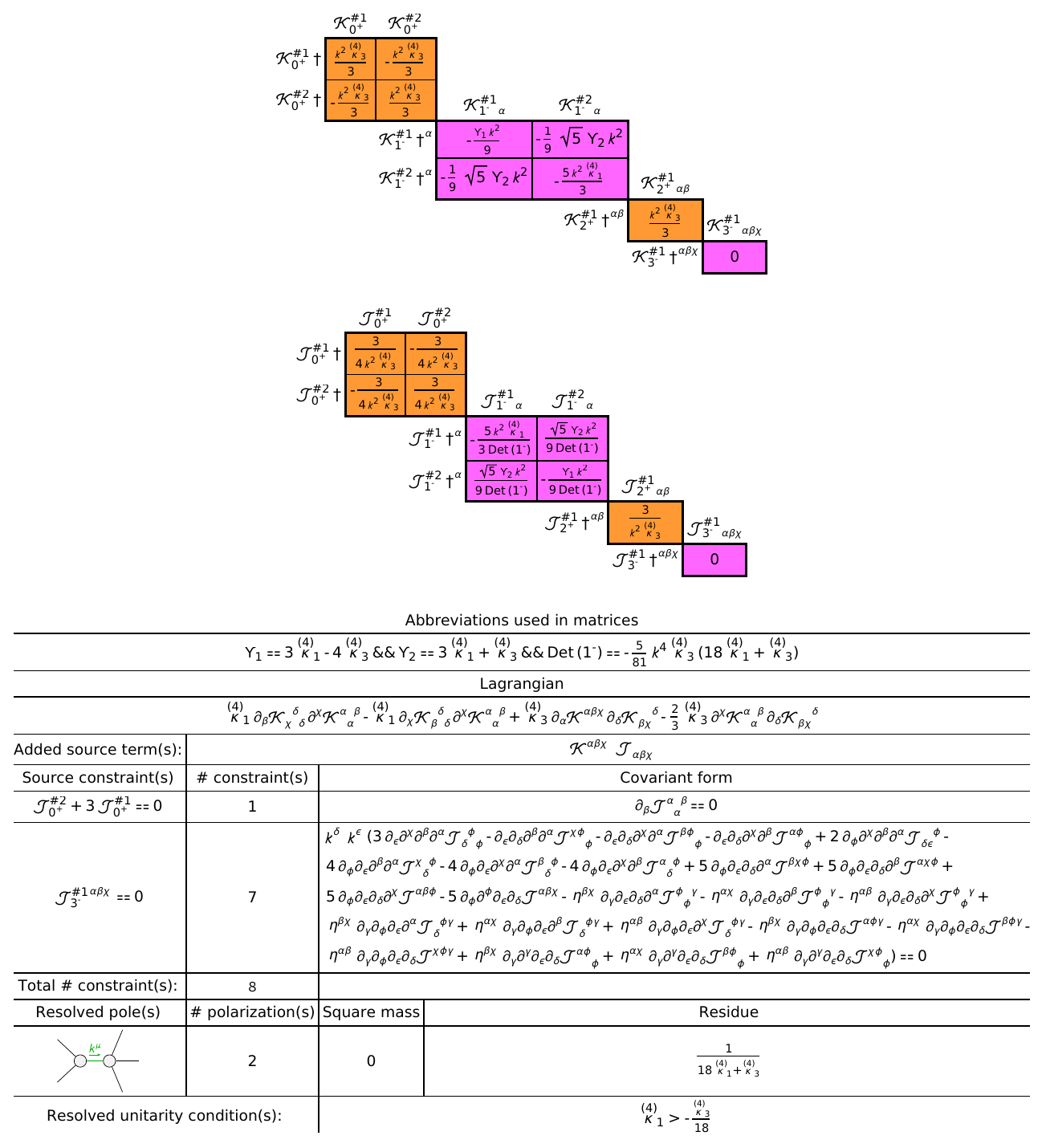}
	\caption{Output generated by \PSALTer{}. The spectrograph of~\NamedModel{S123C1E1}, as defined in~\cref{AllModelsS123}. All notation is defined in~\cref{FieldKinematicsS123Field,ParticleSpectrographS123}. In common with~\NamedModel{S123B1C1D1F2} in~\cref{ParticleSpectrographS123B1C1D1F2} this is a~$J=1$ model.}
\label{ParticleSpectrographS123C1E1}
\end{figure*}

\begin{figure*}[htbp]
	\includegraphics[width=\linewidth]{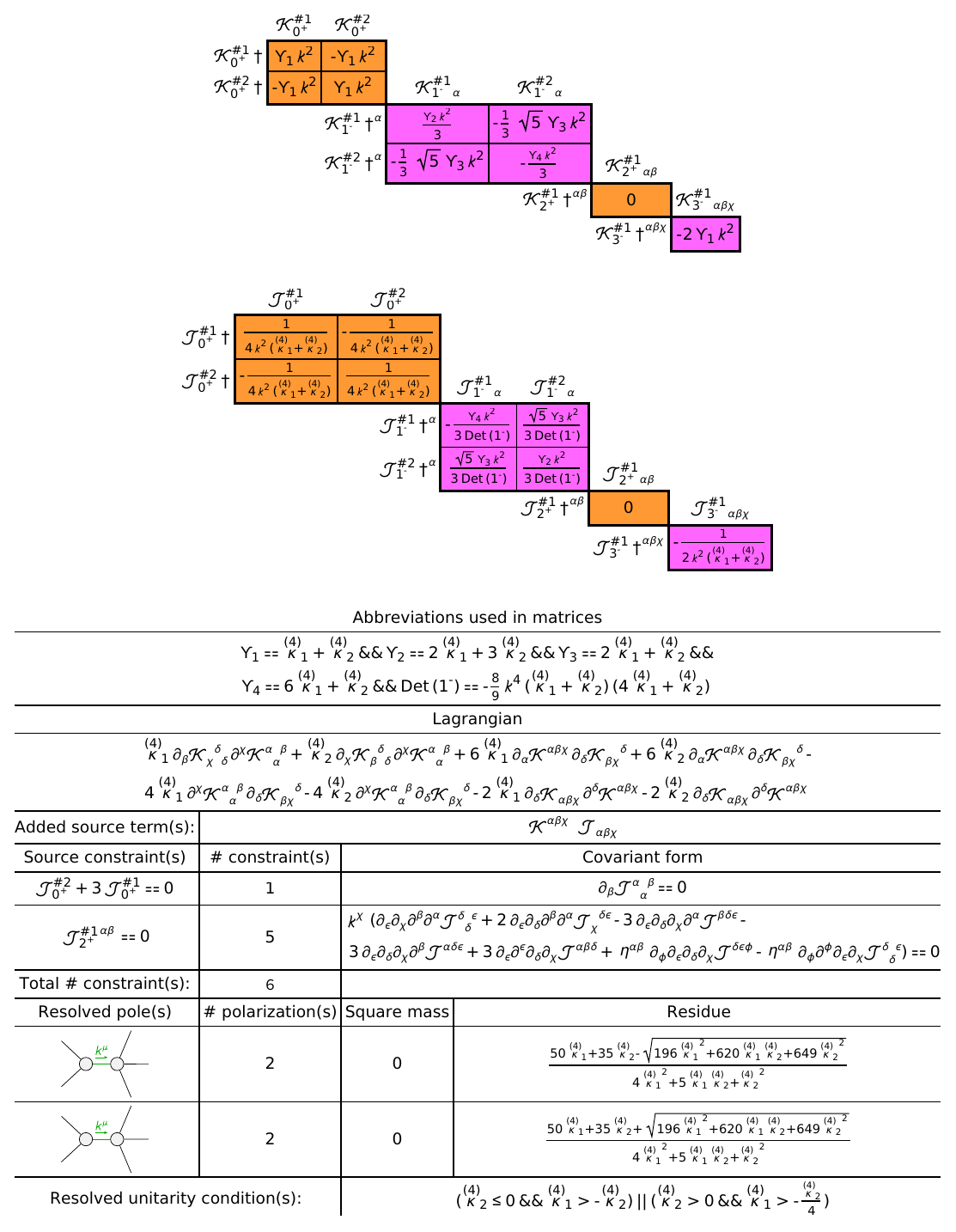}
	\caption{Output generated by \PSALTer{}. The spectrograph of~\NamedModel{S123C1E2}, as defined in~\cref{AllModelsS123}. All notation is defined in~\cref{FieldKinematicsS123Field,ParticleSpectrographS123}. This is a generalisation of the model found in~\cite{Campoleoni:2012th}, to which it reduces under the assumption~$\Dis{^{\alpha}_{\alpha}^{\beta}}\equiv 0$. The model propagates~$J=1$ and~$J=3$ simultaneously.}
\label{ParticleSpectrographS123C1E2}
\end{figure*}

\begin{figure*}[htbp]
	\includegraphics[width=\linewidth]{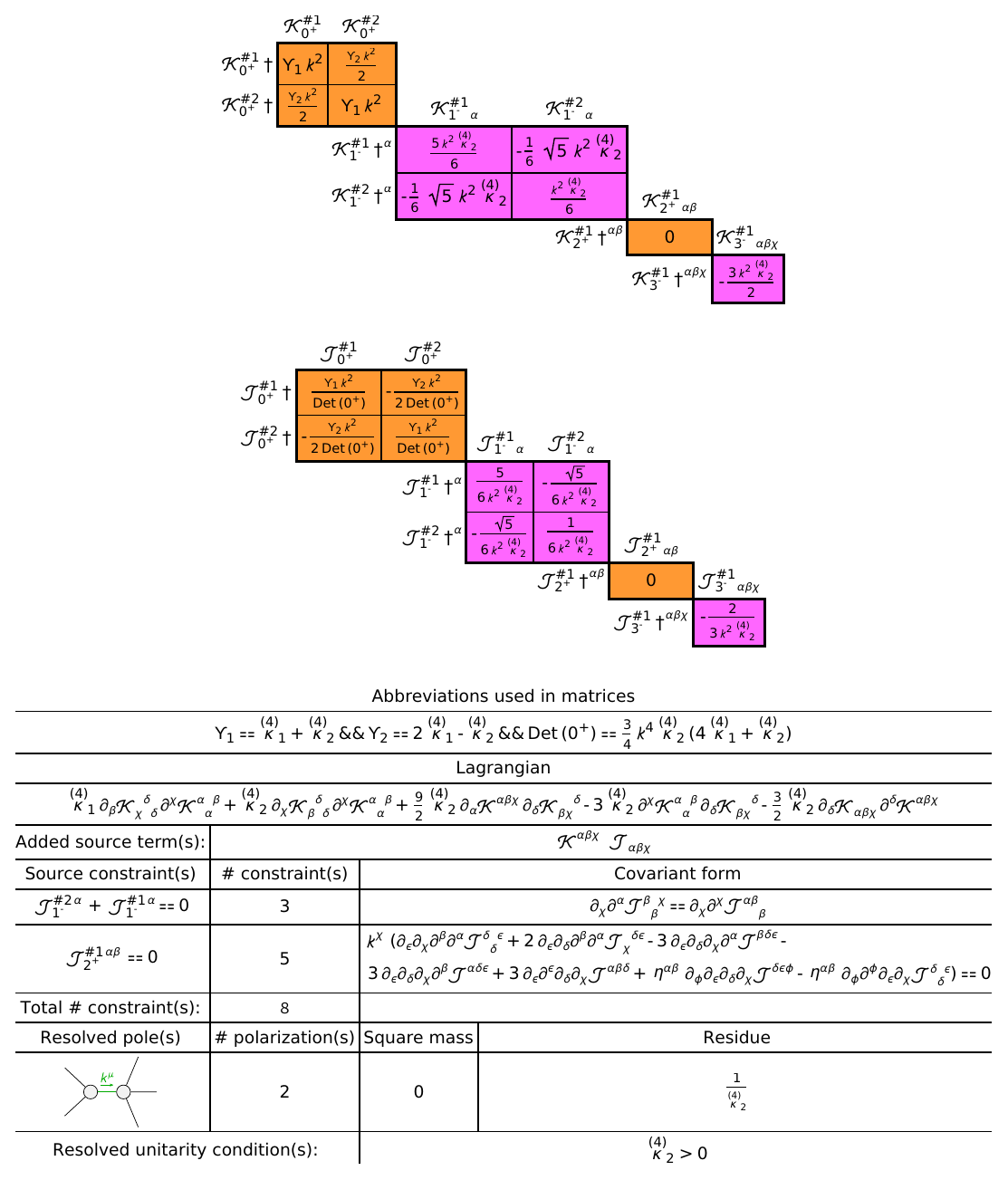}
	\caption{Output generated by \PSALTer{}. The spectrograph of~\NamedModel{S123C2E2}, as defined in~\cref{AllModelsS123}. All notation is defined in~\cref{FieldKinematicsS123Field,ParticleSpectrographS123}. This model is a superposition of the Fronsdal theory~\NamedModel{S123C1D1F2} in~\cref{ParticleSpectrographS123C1D1F2} and the non-propagating model~\NamedModel{S123B1C1D1F1}.}
\label{ParticleSpectrographS123C2E2}
\end{figure*}

\begin{figure*}[htbp]
	\includegraphics[width=\linewidth]{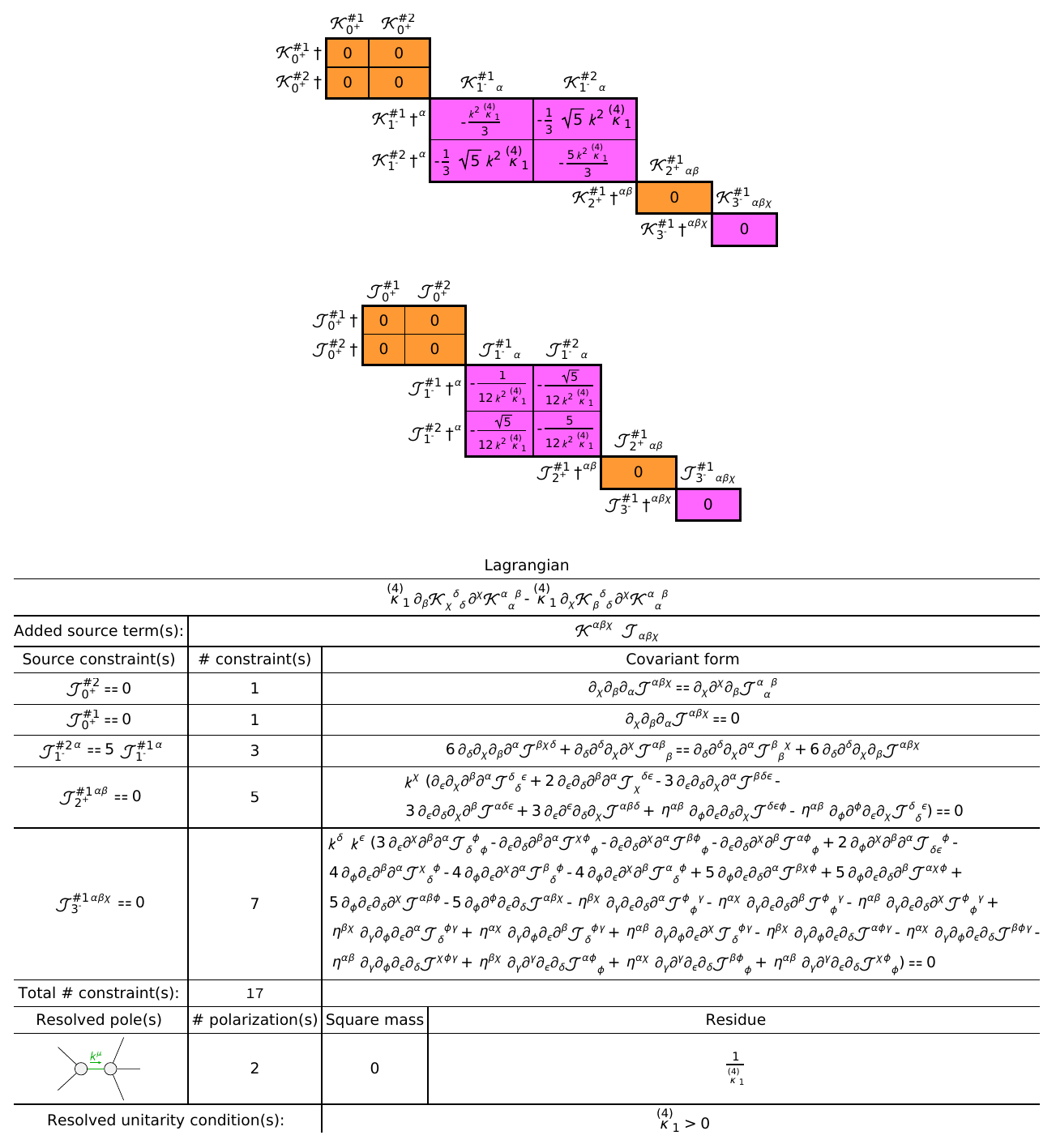}
	\caption{Output generated by \PSALTer{}. The spectrograph of~\NamedModel{S123B1C1D1F2}, as defined in~\cref{AllModelsS123}. All notation is defined in~\cref{FieldKinematicsS123Field,ParticleSpectrographS123}. This Maxwell-type model was found previously in~\cite{Barker:2025xzd}.}
\label{ParticleSpectrographS123B1C1D1F2}
\end{figure*}

\begin{figure*}[htbp]
	\includegraphics[width=\linewidth]{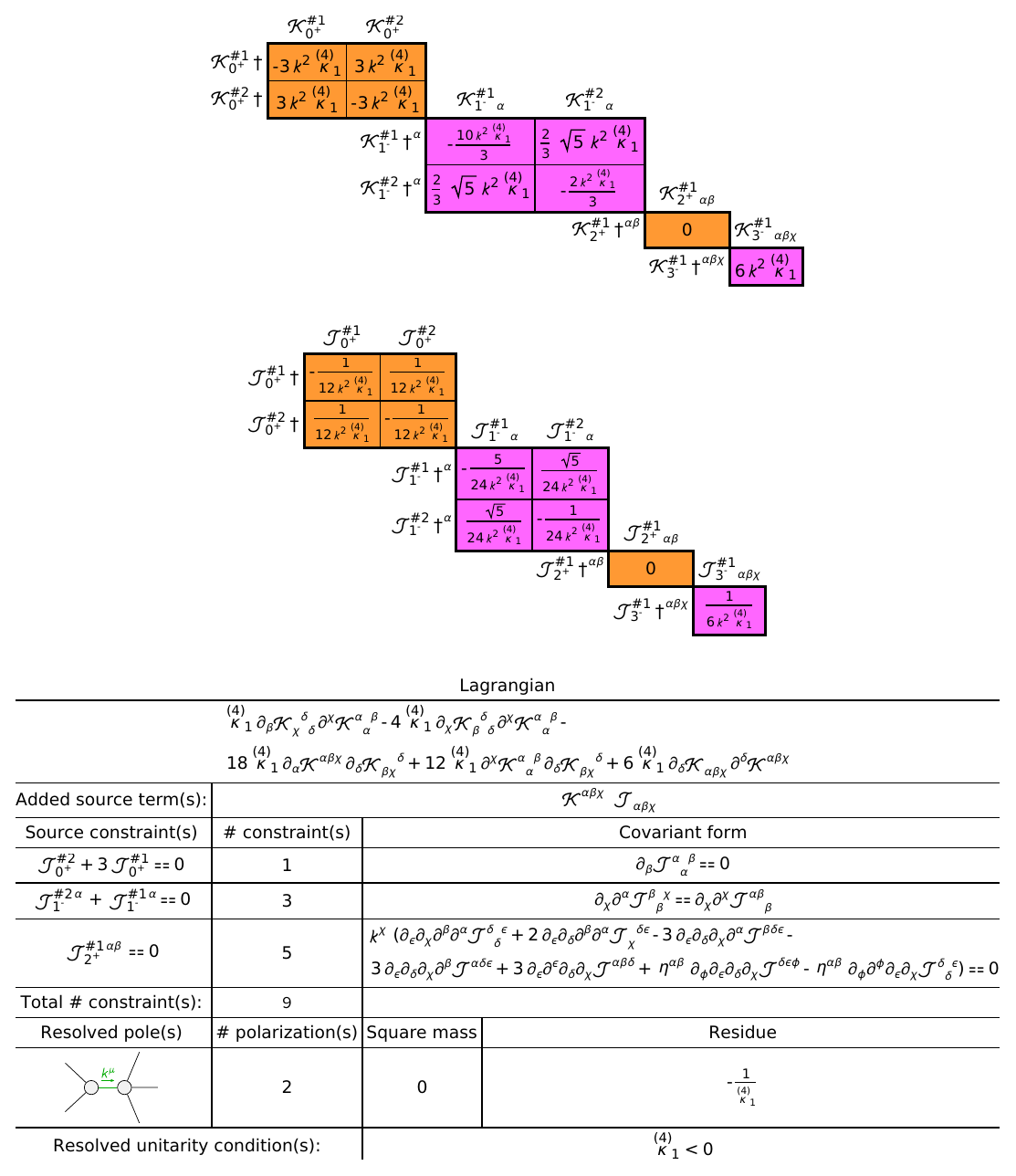}
	\caption{Output generated by \PSALTer{}. The spectrograph of~\NamedModel{S123C1D1F2}, as defined in~\cref{AllModelsS123}. All notation is defined in~\cref{FieldKinematicsS123Field,ParticleSpectrographS123}. This is Fronsdal theory, which propagates the~$J=3$ sector~\cite{Fronsdal:1978rb}.}
\label{ParticleSpectrographS123C1D1F2}
\end{figure*}

\paragraph*{Summary} Using the algorithm presented in~\cref{Sec:Methods} we obtain 23 symmetric models between the root theory~\NamedModel{S123} and the trivial empty theory~\NamedModel{S123Z1}, for which~$\Lagrangian{S123Z1}\equiv 0$. These are illustrated in~\cref{GraphRepresentationS123}. The notation allocates a letter to signify the number of constraints (where~$\mathfrak{B}$ signifies two constraints,~$\mathfrak{C}$ signifies three constraints, and so on) and a subscript to distinguish different models. The full details of these models are provided in~\cref{Sec:Parameters} and~\cref{AllModelsS123}. Among these models, we find five unitary cases~\NamedModel{S123C1E1},~\NamedModel{S123C1E2},~\NamedModel{S123C2E2},~\NamedModel{S123B1C1D1F2}, and~\NamedModel{S123C1D1F2}, whose full spectrographs are provided in~\cref{ParticleSpectrographS123C1E1,ParticleSpectrographS123C1E2,ParticleSpectrographS123C2E2,ParticleSpectrographS123B1C1D1F2,ParticleSpectrographS123C1D1F2}, respectively.

\paragraph*{Spin one} An important consistency check emerges in the identification of model~\NamedModel{S123B1C1D1F2} in~\cref{ParticleSpectrographS123B1C1D1F2}, which confirms our earlier analysis based on a natural ansatz for the defining gauge symmetry as discussed in~\cite{Barker:2025xzd}. In contrast, the newly found model~\NamedModel{S123C1E1} displays an intriguing structure characterised by multiple parameters while remaining the most general form permitted by the given symmetry. The presence of~$k^2$ terms in the~$J=2$ sector of \cref{ParticleSpectrographS123C1E1} might naively suggest particle propagation in that channel. However, a helicity decomposition definitively establishes that only~$J=1$ particles propagate within this model.

\paragraph*{Spin three} A further valuable validation of our methodology is the identification of model~\NamedModel{S123C1D1F2} in~\cref{ParticleSpectrographS123C1D1F2}, which aligns exactly with Fronsdal’s formulation for massless~$J=3$ particle propagation~\cite{Fronsdal:1978rb}. Closely related is model~\NamedModel{S123C2E2} in~\cref{ParticleSpectrographS123C2E2}, initially appearing to contain multiple propagating sectors and coupling constants. Nevertheless, spectral analysis conclusively demonstrates that only two physical states propagate, indicating the involvement of a single physical sector. Furthermore, only one coupling parameter influences the unitarity conditions, suggesting potential redundancy among couplings. A deeper examination confirms this intuition, revealing that this model decomposes precisely into Fronsdal's~$J=3$ structure combined with model~\NamedModel{S123B1C1D1F1}. As established previously, the latter model contributes no physical states to the spectrum (see~\cref{AllModelsS123}).

\paragraph*{Spin one and spin three} Finally, we identify model~\NamedModel{S123C1E2} as the unique instance of a symmetric framework capable of simultaneous propagation, as illustrated in \cref{ParticleSpectrographS123C1E2}. To our knowledge, this particular characteristic has previously appeared only within the Maxwell-like theories presented by Campoleoni and Francia in~\cite{Campoleoni:2012th}. A key prerequisite for achieving their model is the adoption of traceless fields, a significant distinction from our unconstrained approach in~\cref{eq:DisS123}. Imposition of the traceless condition~$\Dis{^{\alpha}_{\alpha}^{\beta}}\equiv 0$ on~$\Lagrangian{S123C1E2}$ establishes an equivalence with their approach through a straightforward field redefinition. Thus, it is both surprising and noteworthy to observe that simultaneous propagation, typically reliant on traceless tensors, can also be realised using unconstrained tensors and gauge symmetries.

\paragraph*{Further work} The apparatus presented in~\cref{Algorithm} can be applied to any class of models accepted by \PSALTer{}, i.e. quadratic field theories involving arbitrary numbers of arbitrarily symmetrized tensor fields up to rank three, with or without parity violation. Since \PSALTer is parallelised, the algorithm may also be scaled using high-performance computing resources. Regarding the current science products, we caution that unitarity of the most general model for a given gauge symmetry is only a necessary condition for the non-linear completion of the model in~\cref{EFTLag} to be found. Indeed, consistent completion of the symmetry at higher orders is already known to be problematic for the Fronsdal case~\cite{Boulanger:2000rq,Bekaert:2010hp}. Finally, notice that, since our algorithm only deals with conditions linear in the coefficients, symmetric models deriving from quadratic deconfliction constraints are not considered here. However, such special cases are only relevant for the reduction steps concerning the models \NamedModel{S123B1}, \NamedModel{S123B2} and \NamedModel{S123} itself.

\begin{acknowledgments}
This work was improved by useful discussions with Dario Francia, and Roberto Percacci.

This work used the DiRAC Data Intensive service~(CSD3 \href{www.csd3.cam.ac.uk}{www.csd3.cam.ac.uk}) at the University of Cambridge, managed by the University of Cambridge University Information Services on behalf of the STFC DiRAC HPC Facility~(\href{www.dirac.ac.uk}{www.dirac.ac.uk}). The DiRAC component of CSD3 at Cambridge was funded by BEIS, UKRI and STFC capital funding and STFC operations grants. DiRAC is part of the UKRI Digital Research Infrastructure.

This work also used the Newton compute server, access to which was provisioned by Will Handley using an ERC grant.

W.~B. is grateful for the support of Girton College, Cambridge, Marie Skłodowska-Curie Actions and the Institute of Physics of the Czech Academy of Sciences. The work of C.~M. was supported by the Estonian Research Council grant PRG1677. A.~S. acknowledges financial support from the ANID CONICYT-PFCHA/DoctoradoNacional/2020-21201387.

Co-funded by the European Union (Physics for Future – Grant Agreement No. 101081515). Views and opinions expressed are however those of the author(s) only and do not necessarily reflect those of the European Union or European Research Executive Agency. Neither the European Union nor the granting authority can be held responsible for them.
\end{acknowledgments}

\bibliography{Manuscript}

\appendix

\section{Catalogue}\label{Sec:Parameters}

In this appendix we provide~\cref{AllModelsS123}, which contains full definitions of all the models in~\cref{GraphRepresentationS123}.

\def\arraystretch{1.5}
\begin{longtable*}{ @{\extracolsep{0pt}} >{\centering\arraybackslash}m{0.07\linewidth} | >{\centering\arraybackslash}m{0.43\linewidth} | >{\centering\arraybackslash}m{0.43\linewidth} | >{\centering\arraybackslash}m{0.07\linewidth}}
\caption{Complete definitions of all models in~\cref{GraphRepresentationS123}. The second column lists all linear combinations of the parameters in~\cref{RootTheoryS123} which must vanish in order to reach the model in question. The third column lists all linear combinations of the parameters in~\cref{RootTheoryS123} which lead to further special cases. The spectrum is inconsistent (\SampleInconsistent{}), consistent (\SampleConsistent{}) or empty (\SampleEmpty{}). The consistent spectra are illustrated in~\cref{ParticleSpectrographS123C1E1,ParticleSpectrographS123C1E2,ParticleSpectrographS123C2E2,ParticleSpectrographS123B1C1D1F2,ParticleSpectrographS123C1D1F2}.}\label{AllModelsS123}\\
\hline\hline
Model & Definition & Deconfliction & Unitary \\
\hline
\endfirsthead
\multicolumn{4}{c}{Table \thetable{} -- continued from previous page} \\
\hline\hline
\endhead
\hline\hline
\multicolumn{4}{r}{Continued on next page}
\endfoot
\hline\hline
\endlastfoot
\NamedModel{S123} &~--- &~$\Big[ \MAGG{4}{_{7}}=0 \wedge \MAGG{2}{_{1}}=0 \Big] \vee \Big[ \MAGG{4}{_{3}} + 3\MAGG{4}{_{7}}=0 \wedge \MAGG{2}{_{1}}=0 \Big] \vee \Big[ \MAGG{4}{_{3}} + \tfrac{3}{2}\MAGG{4}{_{4}}=0 \wedge 2\MAGG{4}{_{1}} + 2\MAGG{4}{_{2}} + \MAGG{4}{_{7}}=0 \wedge \MAGG{2}{_{1}} + 2\MAGG{2}{_{2}}=0 \Big] \vee \Big[ \MAGG{4}{_{3}} + \tfrac{3}{2}\MAGG{4}{_{4}}=0 \wedge 12\MAGG{4}{_{2}} + \MAGG{4}{_{4}} + 6\MAGG{4}{_{7}}=0 \wedge \MAGG{2}{_{1}} + 2\MAGG{2}{_{2}}=0 \Big] \vee \Big[ 5\MAGG{4}{_{4}} - 6\MAGG{4}{_{7}}=0 \wedge \MAGG{4}{_{3}} + \tfrac{9}{5}\MAGG{4}{_{7}}=0 \wedge \MAGG{2}{_{1}}=0 \Big] \vee \Big[ \MAGG{4}{_{4}} - \MAGG{4}{_{7}}=0 \wedge \MAGG{4}{_{3}} + \tfrac{3}{2}\MAGG{4}{_{7}}=0 \wedge \MAGG{2}{_{1}}=0 \Big]$ & \Inconsistent{} \\
\hline
\NamedModel{S123B1} &~$\MAGG{4}{_{7}}=0 \wedge \MAGG{2}{_{1}}=0$ &~$\MAGG{4}{_{3}}=0$ & \Inconsistent{} \\
\hline
\NamedModel{S123B2} &~$\tfrac{1}{3}\MAGG{4}{_{3}} + \MAGG{4}{_{7}}=0 \wedge \MAGG{2}{_{1}}=0$ &~$\MAGG{4}{_{3}}=0$ & \Inconsistent{} \\
\hline
\NamedModel{S123C1} &~$\tfrac{2}{3}\MAGG{4}{_{3}} + \MAGG{4}{_{4}}=0 \wedge 2\MAGG{4}{_{1}} + 2\MAGG{4}{_{2}} + \MAGG{4}{_{7}}=0 \wedge \tfrac{1}{2}\MAGG{2}{_{1}} + \MAGG{2}{_{2}}=0$ &~$18\MAGG{4}{_{1}} + \MAGG{4}{_{3}}=0 \vee \Big[ \MAGG{4}{_{1}} + \MAGG{4}{_{2}}=0 \wedge \MAGG{2}{_{1}}=0 \Big] \vee \Big[ 6\MAGG{4}{_{1}} + 6\MAGG{4}{_{2}} - \MAGG{4}{_{3}}=0 \wedge \MAGG{2}{_{1}}=0 \Big] \vee \Big[ 18\MAGG{4}{_{1}} + 18\MAGG{4}{_{2}} - 5\MAGG{4}{_{3}}=0 \wedge \MAGG{2}{_{1}}=0 \Big] \vee \Big[ 3\MAGG{4}{_{1}} + 3\MAGG{4}{_{2}} - \MAGG{4}{_{3}}=0 \wedge \MAGG{2}{_{1}}=0 \Big]$ & \Inconsistent{} \\
\hline
\NamedModel{S123C2} &~$\tfrac{2}{3}\MAGG{4}{_{3}} + \MAGG{4}{_{4}}=0 \wedge 2\MAGG{4}{_{2}} - \tfrac{1}{9}\MAGG{4}{_{3}} + \MAGG{4}{_{7}}=0 \wedge \tfrac{1}{2}\MAGG{2}{_{1}} + \MAGG{2}{_{2}}=0$ &~$18\MAGG{4}{_{1}} + \MAGG{4}{_{3}}=0 \vee \Big[ 18\MAGG{4}{_{2}} - \MAGG{4}{_{3}}=0 \wedge \MAGG{2}{_{1}}=0 \Big] \vee \Big[ 9\MAGG{4}{_{2}} - 2\MAGG{4}{_{3}}=0 \wedge \MAGG{2}{_{1}}=0 \Big] \vee \Big[ 3\MAGG{4}{_{2}} - \MAGG{4}{_{3}}=0 \wedge \MAGG{2}{_{1}}=0 \Big] \vee \Big[ 18\MAGG{4}{_{2}} - 7\MAGG{4}{_{3}}=0 \wedge \MAGG{2}{_{1}}=0 \Big]$ & \Inconsistent{} \\
\hline
\NamedModel{S123C3} &~$\tfrac{2}{3}\MAGG{4}{_{3}} + \MAGG{4}{_{4}}=0 \wedge \tfrac{5}{9}\MAGG{4}{_{3}} + \MAGG{4}{_{7}}=0 \wedge \MAGG{2}{_{1}}=0$ &~$\MAGG{4}{_{3}}=0 \vee \Big[ 18\MAGG{4}{_{1}} + 18\MAGG{4}{_{2}} - 5\MAGG{4}{_{3}}=0 \wedge \MAGG{2}{_{2}}=0 \Big] \vee \Big[ 3\MAGG{4}{_{2}} - \MAGG{4}{_{3}}=0 \wedge \MAGG{2}{_{2}}=0 \Big]$ & \Inconsistent{} \\
\hline
\NamedModel{S123C4} &~$\tfrac{2}{3}\MAGG{4}{_{3}} + \MAGG{4}{_{4}}=0 \wedge \tfrac{2}{3}\MAGG{4}{_{3}} + \MAGG{4}{_{7}}=0 \wedge \MAGG{2}{_{1}}=0$ &~$\MAGG{4}{_{3}}=0 \vee \Big[ 3\MAGG{4}{_{1}} + 3\MAGG{4}{_{2}} - \MAGG{4}{_{3}}=0 \wedge \MAGG{2}{_{2}}=0 \Big] \vee \Big[ 18\MAGG{4}{_{2}} - 7\MAGG{4}{_{3}}=0 \wedge \MAGG{2}{_{2}}=0 \Big]$ & \Inconsistent{} \\
\hline
\NamedModel{S123B1C1} &~$\MAGG{4}{_{3}}=0 \wedge \MAGG{4}{_{7}}=0 \wedge \MAGG{2}{_{1}}=0$ &~$\MAGG{4}{_{4}}=0$ & \Empty{} \\
\hline
\NamedModel{S123B1C1D1} &~$\MAGG{4}{_{3}}=0 \wedge \MAGG{4}{_{4}}=0 \wedge \MAGG{4}{_{7}}=0 \wedge \MAGG{2}{_{1}}=0$ &~$ \Big[ \MAGG{4}{_{1}} + \MAGG{4}{_{2}} =0 \wedge \MAGG{2}{_{2}}=0 \Big] \vee \Big[ \MAGG{4}{_{2}}=0 \wedge \MAGG{2}{_{2}}=0 \Big]$ & \Inconsistent{} \\
\hline
\NamedModel{S123C1D1} &~$18\MAGG{4}{_{1}} + \MAGG{4}{_{3}}=0 \wedge -12\MAGG{4}{_{1}} + \MAGG{4}{_{4}}=0 \wedge 2\MAGG{4}{_{1}} + 2\MAGG{4}{_{2}} + \MAGG{4}{_{7}}=0 \wedge \tfrac{1}{2}\MAGG{2}{_{1}} + \MAGG{2}{_{2}}=0$ &~$ \Big[ \MAGG{4}{_{1}} + \MAGG{4}{_{2}}=0 \wedge \MAGG{2}{_{1}}=0 \Big] \vee \Big[ 7\MAGG{4}{_{1}} + \MAGG{4}{_{2}}=0 \wedge \MAGG{2}{_{1}}=0 \Big] \vee \Big[ 6\MAGG{4}{_{1}}  + \MAGG{4}{_{2}}=0 \wedge \MAGG{2}{_{1}}=0 \Big] \vee \Big[ 4\MAGG{4}{_{1}} + \MAGG{4}{_{2}}=0 \wedge \MAGG{2}{_{1}}=0 \Big] $ & \Inconsistent{} \\
\hline
\NamedModel{S123C1E1} &~$\MAGG{4}{_{1}} + \MAGG{4}{_{2}}=0 \wedge \tfrac{2}{3}\MAGG{4}{_{3}} + \MAGG{4}{_{4}}=0 \wedge \MAGG{4}{_{7}}=0 \wedge \MAGG{2}{_{1}}=0 \wedge \MAGG{2}{_{2}}=0$ &~$\MAGG{4}{_{3}}=0 \vee 18\MAGG{4}{_{1}} + \MAGG{4}{_{3}}=0$ & \Consistent{} \\
\hline
\NamedModel{S123C1E2} &~$-6\MAGG{4}{_{1}} - 6\MAGG{4}{_{2}} + \MAGG{4}{_{3}}=0 \wedge 4\MAGG{4}{_{1}} + 4\MAGG{4}{_{2}} + \MAGG{4}{_{4}}=0 \wedge 2\MAGG{4}{_{1}} + 2\MAGG{4}{_{2}} + \MAGG{4}{_{7}}=0 \wedge \MAGG{2}{_{1}}=0 \wedge \MAGG{2}{_{2}}=0$ &~$\MAGG{4}{_{1}} + \MAGG{4}{_{2}}=0 \vee 4\MAGG{4}{_{1}} + \MAGG{4}{_{2}}=0$ & \Consistent{} \\
\hline
\NamedModel{S123C1E3} &~$-\tfrac{18}{5}\MAGG{4}{_{1}} - \tfrac{18}{5}\MAGG{4}{_{2}} + \MAGG{4}{_{3}}=0 \wedge \tfrac{12}{5}\MAGG{4}{_{1}} + \tfrac{12}{5}\MAGG{4}{_{2}} + \MAGG{4}{_{4}}=0 \wedge 2\MAGG{4}{_{1}} + 2\MAGG{4}{_{2}} + \MAGG{4}{_{7}}=0 \wedge \MAGG{2}{_{1}}=0 \wedge \MAGG{2}{_{2}}=0$ &~$\MAGG{4}{_{1}} + \MAGG{4}{_{2}}=0 \vee 6\MAGG{4}{_{1}} + \MAGG{4}{_{2}}=0$ & \Inconsistent{} \\
\hline
\NamedModel{S123C1E4} &~$-3\MAGG{4}{_{1}} - 3\MAGG{4}{_{2}} + \MAGG{4}{_{3}}=0 \wedge 2\MAGG{4}{_{1}} + 2\MAGG{4}{_{2}} + \MAGG{4}{_{4}}=0 \wedge 2\MAGG{4}{_{1}} + 2\MAGG{4}{_{2}} + \MAGG{4}{_{7}}=0 \wedge \MAGG{2}{_{1}}=0 \wedge \MAGG{2}{_{2}}=0$ &~$\MAGG{4}{_{1}} + \MAGG{4}{_{2}}=0 \vee 7\MAGG{4}{_{1}} + \MAGG{4}{_{2}}=0$ & \Inconsistent{} \\
\hline
\NamedModel{S123C2E1} &~$-18\MAGG{4}{_{2}} + \MAGG{4}{_{3}}=0 \wedge 12\MAGG{4}{_{2}} + \MAGG{4}{_{4}}=0 \wedge \MAGG{4}{_{7}}=0 \wedge \MAGG{2}{_{1}}=0 \wedge \MAGG{2}{_{2}}=0$ &~$\MAGG{4}{_{2}}=0 \vee \MAGG{4}{_{1}} + \MAGG{4}{_{2}}=0$ & \Empty{} \\
\hline
\NamedModel{S123C2E2} &~$-\tfrac{9}{2}\MAGG{4}{_{2}} + \MAGG{4}{_{3}}=0 \wedge 3\MAGG{4}{_{2}} + \MAGG{4}{_{4}}=0 \wedge \tfrac{3}{2}\MAGG{4}{_{2}} + \MAGG{4}{_{7}}=0 \wedge \MAGG{2}{_{1}}=0 \wedge \MAGG{2}{_{2}}=0$ &~$\MAGG{4}{_{2}}=0 \vee 4\MAGG{4}{_{1}} + \MAGG{4}{_{2}}=0$ & \Consistent{} \\
\hline
\NamedModel{S123C2E3} &~$-3\MAGG{4}{_{2}} + \MAGG{4}{_{3}}=0 \wedge 2\MAGG{4}{_{2}} + \MAGG{4}{_{4}}=0 \wedge \tfrac{5}{3}\MAGG{4}{_{2}} + \MAGG{4}{_{7}}=0 \wedge \MAGG{2}{_{1}}=0 \wedge \MAGG{2}{_{2}}=0$ &~$\MAGG{4}{_{2}}=0 \vee 6\MAGG{4}{_{1}} + \MAGG{4}{_{2}}=0$ & \Inconsistent{} \\
\hline
\NamedModel{S123C2E4} &~$-\tfrac{18}{7}\MAGG{4}{_{2}} + \MAGG{4}{_{3}}=0 \wedge \tfrac{12}{7}\MAGG{4}{_{2}} + \MAGG{4}{_{4}}=0 \wedge \tfrac{12}{7}\MAGG{4}{_{2}} + \MAGG{4}{_{7}}=0 \wedge \MAGG{2}{_{1}}=0 \wedge \MAGG{2}{_{2}}=0$ &~$\MAGG{4}{_{2}}=0 \vee 7\MAGG{4}{_{1}} + \MAGG{4}{_{2}}=0$ & \Inconsistent{} \\
\hline
\NamedModel{S123B1C1D1F1} &~$\MAGG{4}{_{2}}=0 \wedge \MAGG{4}{_{3}}=0 \wedge \MAGG{4}{_{4}}=0 \wedge \MAGG{4}{_{7}}=0 \wedge \MAGG{2}{_{1}}=0 \wedge \MAGG{2}{_{2}}=0$ &~$\MAGG{4}{_{1}}=0$ & \Empty{} \\
\hline
\NamedModel{S123B1C1D1F2} &~$\MAGG{4}{_{1}} + \MAGG{4}{_{2}}=0 \wedge \MAGG{4}{_{3}}=0 \wedge \MAGG{4}{_{4}}=0 \wedge \MAGG{4}{_{7}}=0 \wedge \MAGG{2}{_{1}}=0 \wedge \MAGG{2}{_{2}}=0$ &~$\MAGG{4}{_{1}}=0$ & \Consistent{} \\
\hline
\NamedModel{S123C1D1F1} &~$\MAGG{4}{_{1}} + \MAGG{4}{_{2}}=0 \wedge 18\MAGG{4}{_{1}} + \MAGG{4}{_{3}}=0 \wedge -12\MAGG{4}{_{1}} + \MAGG{4}{_{4}}=0 \wedge \MAGG{4}{_{7}}=0 \wedge \MAGG{2}{_{1}}=0 \wedge \MAGG{2}{_{2}}=0$ &~$\MAGG{4}{_{1}}=0$ & \Empty{} \\
\hline
\NamedModel{S123C1D1F2} &~$4\MAGG{4}{_{1}} + \MAGG{4}{_{2}}=0 \wedge 18\MAGG{4}{_{1}} + \MAGG{4}{_{3}}=0 \wedge -12\MAGG{4}{_{1}} + \MAGG{4}{_{4}}=0 \wedge -6\MAGG{4}{_{1}} + \MAGG{4}{_{7}}=0 \wedge \MAGG{2}{_{1}}=0 \wedge \MAGG{2}{_{2}}=0$ &~$\MAGG{4}{_{1}}=0$ & \Consistent{} \\
\hline
\NamedModel{S123C1D1F3} &~$6\MAGG{4}{_{1}} + \MAGG{4}{_{2}}=0 \wedge 18\MAGG{4}{_{1}} + \MAGG{4}{_{3}}=0 \wedge -12\MAGG{4}{_{1}} + \MAGG{4}{_{4}}=0 \wedge -10\MAGG{4}{_{1}} + \MAGG{4}{_{7}}=0 \wedge \MAGG{2}{_{1}}=0 \wedge \MAGG{2}{_{2}}=0$ &~$\MAGG{4}{_{1}}=0$ & \Inconsistent{} \\
\hline
\NamedModel{S123C1D1F4} &~$7\MAGG{4}{_{1}} + \MAGG{4}{_{2}}=0 \wedge 18\MAGG{4}{_{1}} + \MAGG{4}{_{3}}=0 \wedge -12\MAGG{4}{_{1}} + \MAGG{4}{_{4}}=0 \wedge -12\MAGG{4}{_{1}} + \MAGG{4}{_{7}}=0 \wedge \MAGG{2}{_{1}}=0 \wedge \MAGG{2}{_{2}}=0$ &~$\MAGG{4}{_{1}}=0$ & \Inconsistent{} \\
\hline
\NamedModel{S123Z1} &~$\MAGG{4}{_{1}}=0 \wedge \MAGG{4}{_{2}}=0 \wedge \MAGG{4}{_{3}}=0 \wedge \MAGG{4}{_{4}}=0 \wedge \MAGG{4}{_{7}}=0 \wedge \MAGG{2}{_{1}}=0 \wedge \MAGG{2}{_{2}}=0$ &~--- & \Empty{}

\end{longtable*}

\clearpage
\end{document}